\documentclass[12pt]{iopart}
\usepackage{amsmath,amsfonts,amssymb}
\usepackage{harvard}
\usepackage{bm}
\usepackage{graphicx}

\def\bxi {{\bm \xi}}
\def\bx {{\bf x}}

\def\dint {\int\int}

\usepackage[english]{babel}

\begin{document}

\title[]{High-frequency homogenization of zero frequency stop band
  photonic and phononic crystals}
\author{T. Antonakakis$^{1,2}$, R.~V. Craster$^1$ and S. Guenneau$^3$}
 \address{$^1$ Department of Mathematics, Imperial College London,
   London SW7 2AZ, UK  \\
$^2$ European Organization for Nuclear Research, CERN CH-1211, Geneva 23,
Switzerland \\
$^3$Aix Marseille Université, CNRS, Ecole Centrale Marseille, Institut
Fresnel, 13013 Marseille, France 
}

\begin{abstract}
We present an accurate methodology for representing the physics of waves, for periodic
structures, through 
 effective properties for a replacement bulk medium: This is valid even for media
 with zero
 frequency stop-bands and where high frequency phenomena dominate. 

Since the work of Lord Rayleigh in 1892, low
frequency (or quasi-static) behaviour has been neatly encapsulated in 
effective 
anisotropic media; the various parameters come from asymptotic analysis 
relying upon the ratio of the array pitch to the wavelength being sufficiently
small. However such classical homogenization theories break down in the
high-frequency, or stop band, regime whereby the wavelength to pitch
ratio is of order one. Furthermore, arrays of inclusions with Dirichlet data
lead to a zero frequency stop band, with the salient consequence that classical
homogenization is invalid. 

 Higher frequency phenomena are of significant 
importance in photonics (transverse magnetic waves propagating in infinite conducting
parallel fibers), phononics (anti-plane shear waves propagating in isotropic elastic
materials with inclusions), and platonics (flexural waves
propagating in thin-elastic plates with holes). Fortunately,
the recently proposed high-frequency homogenization (HFH) theory is
 only constrained by the knowledge of standing waves in order to
asymptotically reconstruct dispersion curves and associated
Floquet-Bloch eigenfields: It is capable of accurately representing
zero-frequency stop band structures. The homogenized equations are 
partial differential equations with a dispersive anisotropic homogenized tensor 
 that characterizes the effective medium.

We apply HFH to metamaterials, exploiting
the subtle features of Bloch dispersion curves such as Dirac-like cones,
as well as zero and negative group velocity near stop bands in order
to achieve exciting physical phenomena such as cloaking, lensing
and endoscope effects. These are simulated numerically using finite
elements and compared to
predictions from HFH. An extension of HFH to periodic supercells
enabling complete reconstruction of dispersion curves through an unfolding technique is also introduced.


\end{abstract}

\section{Introduction}
\label{sec:intro}

Over the past 25 years, many significant advances have created a
deep understanding of the optical properties of photonic crystals
(PCs) \cite{yablonovitch87a,john87a}; such periodic structures
prohibit the propagation of light, or allow it only in certain
directions at certain frequencies, or localize light in specified
areas.  This sort of metamaterial (using the consensual terminology
for artificial materials engineered to have desired properties that
may not be found in nature, such as negative refraction, see
e.g. \cite{smith04a,sar_rpp05}) enables a marked enhancement of
 control over light propagation; this arises from, for instance, the
periodic patterning of small metallic inclusions embedded within a
dielectric matrix \cite{pendry96a,nicorovici95a}. PCs are periodic
devices whose spectrum is characterized by photonic band gaps and pass
bands, just as electronic band gaps exist in semiconductors: In PCs,
light propagation is disallowed for certain frequencies in certain
directions.  This effect is well known \cite{joannopoulos08a,zolla05a} and forms the basis of many
devices, including Bragg mirrors, dielectric Fabry-Perot filters, and
distributed feedback lasers; all of these contain low-loss dielectrics
 periodic in one dimension, so are 
one-dimensional PCs. Such mirrors are
tremendously useful, but their reflecting properties critically depend
upon the frequency of the incident wave in regard with its
incidence \cite{markos08a}. For broad frequency ranges, one wishes to reflect
light of any polarization at any angle (which requires a complete
photonic band gap) and for Dirichlet media (i.e. those composed with
microstructure where the field is zero on the microparticles) such a
gap 
occurs at zero-frequency. It is therefore possible to create seismic
shields for low-frequency waves \cite{brule13a,antonakakis13b}. 
Extending these ideas to higher dimensions allows for a much greater range of
optical effects: all-angle negative refraction \cite{luo02a,zengerle87a},
ultra-refraction \cite{farhat10b,craster11a} and cloaking at Dirac-like
cones \cite{chan12a}: We will demonstrate here that an effective
medium can be created that reproduces these effects. 

 The considerable recent
activity in this area is, in part, fuelled by advances in numerical
techniques, based on Fourier expansions in the vector electromagnetic
Maxwell equations \cite{joannopoulos08a}, Finite Elements
\cite{zolla05a} and Multipole and lattice sums \cite{movchan02a} for cylinders, to
name but a few, that allow one to visualize the various effects and design
PCs.  The multipole method takes its root in a seminal
paper of Lord Rayleigh \cite{rayleigh92a} which is the  
foundation stone upon which the current edifice of homogenization
theories is built. More precisely, in that 
paper, John William Strutt, the third Lord Rayleigh, solved 
Laplace's equation in two dimensions for rectangular arrays of
cylinders, and in three dimensions for cubic lattices of
spheres. However, a limitation of Rayleigh's algorithm is that it does not
hold when the volume fraction of inclusions increases. 
Multipole methods, in conjunction with lattice sums,
 overcome such obstacles and lead to the
Rayleigh system which is an infinite linear algebraic system; this 
formulation, in terms of an eigenvalue problem, facilitates the
construction of  dispersion curves and the study of both photonic
and phononic band-gap structures.  In the limit of small 
inclusion radii, when 
propagating modes are very close to plane waves, one can truncate the
Rayleigh system ignoring the effect of higher multipoles, 
 to produce a series of approximations each successively more
 accurate to higher values of filling fraction. 
At the dipole order, one is
able to fit the acoustic band, which has a linear behaviour in the
neighbourhood of zero frequency, except of course in the singular case whereby
Dirichlet data is enforced on the boundary inclusions
\cite{poulton01a} and there is a zero frequency stop-band: This is a
substantial limitation of this truncation. We will derive here
the exact solution for zero radius Dirichlet holes, avoiding multipole
expansions, and show dispersion curves whose features lead to 
interesting topical effects, such as Dirac-like cone cloaking and
degenerate points \cite{chan12a} which emerge very naturally in this
 exact solution. 

\begin{figure}
  \begin{center}
    \includegraphics[scale=0.16]{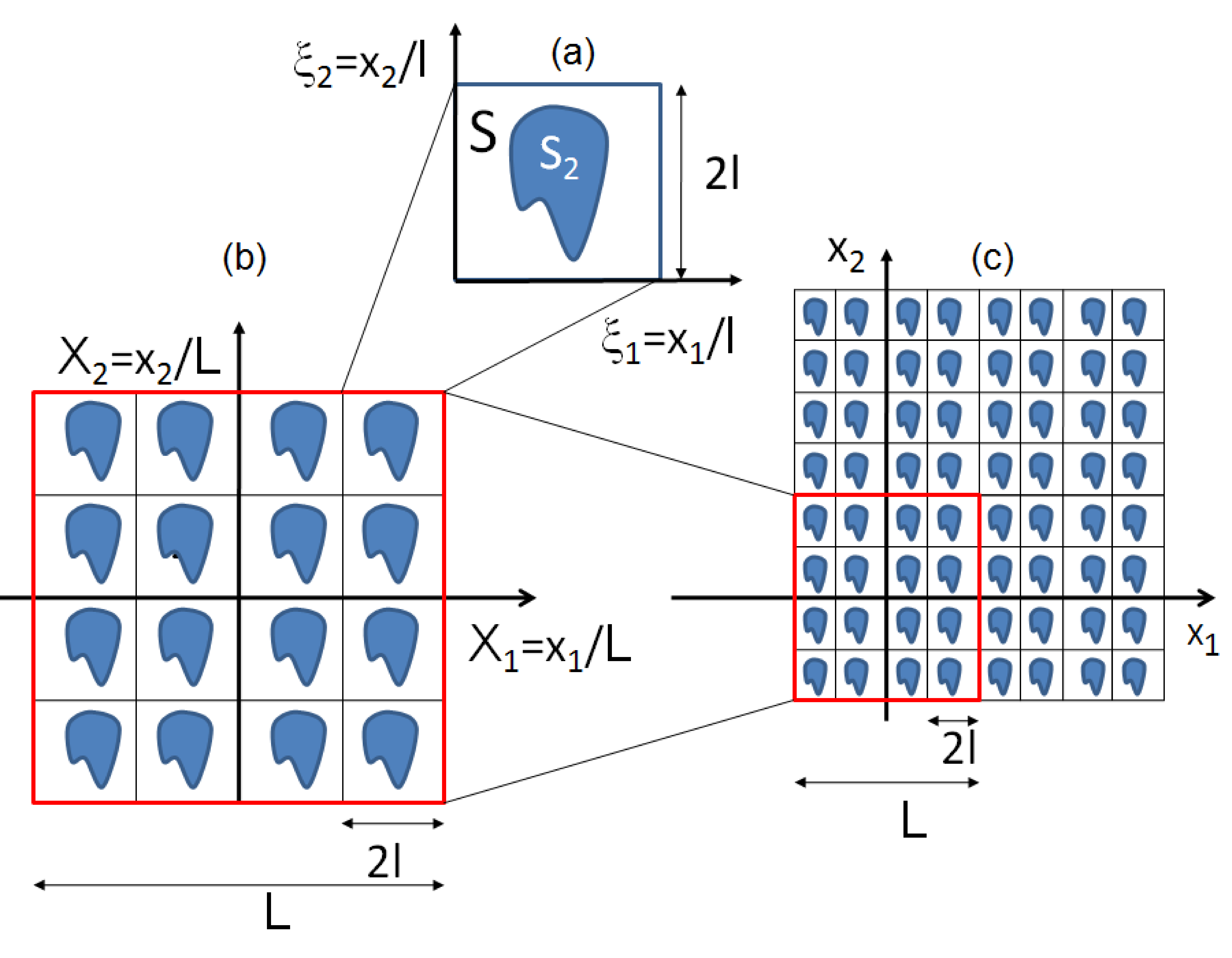} 
   \end{center}
\caption{High Frequency Homogenization (HFH) principle: An elementary cell (a) of sidelength $2l$, modelled
by a fast oscillating variable $\bxi$, is repeated periodically within a
supercell (b) of sidelength $L$, modelled by a slow variable ${\bf X}$, which is itself
repeated periodically in space (c). One then assumes that the parameter
$\epsilon=l/L$ is small, and its vanishing limit
thereafter studied.
The leading order, homogenized, term
of Floquet-Bloch eigenfields within the crystal are then sought as
$u_0({\bf X},\bxi)=f_0({\bf X})U_0(\bxi;\Omega_0)$, wherein $f_0$
accounts for variations of the fields on the order of the
supercells, and $U_0$ captures their fast oscillations in
the much smaller cells, when either periodic or anti-periodic conditions are enforced
on the cells:
Perturbing away from these standing waves of frequency $\Omega_0$
allows for a complete reconstruction of the Bloch spectrum and
associated Floquet-Bloch eigenfields.  
}
\label{fig:homo}
\end{figure} 

Although the numerical approaches discussed briefly are efficient they
still do require substantial computational effort and can obscure
physical understanding and interpretation. 
Here we substantially reduce the numerical complexity of the
wave problem using asymptotic analysis; this has been 
developed over the past 35 years by applied mathematicians primarily for
solving partial differential equations, with rapidly oscillating
periodic coefficients, in the context of thermostatics, continuum
mechanics or electrostatics \cite{bensoussan78a,jikov94a,milton02a}.
The available literature on such effective medium theories is vast, but it
seems that only a very few groups have addressed such problems as the
homogenization of media, with moderate contrast in the material
properties, in three-dimensions \cite{zolla00a,wellander03}
and high-contrast two-dimensional
\cite{zhikov00a,bouchitte04a,cherednichenko06a} photonic crystals,
that have important potential applications in photonics. 
 Besides this,
the aforementioned literature does not address the challenging problem
of homogenization for moderate contrast photonic crystals near stop
band frequencies: Classical homogenization is constrained to low
frequencies and long waves in moderate contrast PCs \cite{silveirinha05a} and so-called high-contrast
homogenization only captures the essence of stop bands in PCs when the
permittivity inside the inclusions is much higher than that of the
surrounding matrix (the contrast being typically on the order to
$\epsilon^{-2}$ where $\epsilon$ is the array pitch which in
turn is
much smaller than the wavelength). This latter area of homogenization
theory is fuelled by interest in artificial magnetism, initiated by
the work of O'Brien and Pendry \cite{obrien02a}. However, moderate
contrast one-dimensional photonic crystals have been recently shown to
display not only artificial magnetism, but also chirality
\cite{yan2013a}, which is an  onset to negative refraction \cite{pendry04b} . 

For all these reasons, there is a strong demand for high-frequency
homogenization theories of PCs in order to grasp, and fully exploit,
the rich behaviour of photonic band gap structures \cite{notomi00,paul11}. This has created a
suite of extended homogenization theories for periodic media called Bloch homogenisation
\cite{conca95a,allaire05a,birman06a,hoefer11a}. There is also a flourishing literature on developing
homogenized elastic media, with frequency dependent effective
parameters, also based upon periodic media \cite{nematnasser11a}: 
 There is considerable interest in creating effective continuum models
 of microstructured media that break free from the conventional low
 frequency homogenisation limitations.

 In this 
paper, we apply the HFH theory proposed by one of us three years ago
\cite{craster10a} to the physics of zero frequency band gap
structures, not only in the context of photonics, but also phononics
for anti-plane shear waves in periodic arrays of inclusions, and
platonics with flexural waves in pinned plates. The latter analysis is
made possible thanks to the extension of HFH to plate theory,
developed by two of us two years ago \cite{antonakakis12a}. Our aim is
to show the universal features of stop band structures thanks to HFH,
and to further exemplify their potential use in control of light and
mechanical waves, with novel applications ranging from cloaking
\cite{milton06a,guenneau07b} to anti-earthquake seismic shields
\cite{antonakakis13a,brule13a}.

\begin{figure}
  \begin{center}
    \includegraphics[scale=0.2]{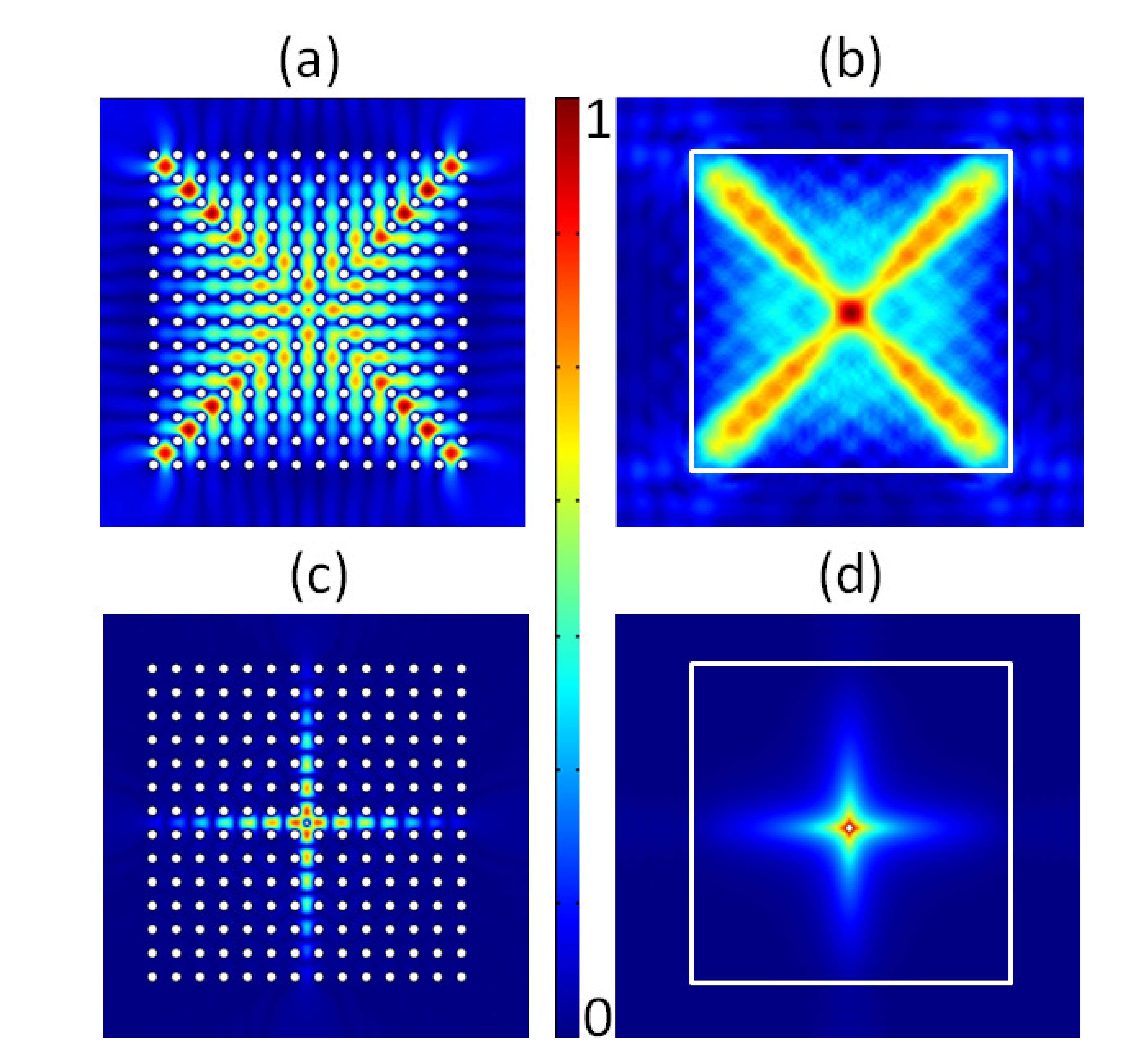} 
   \end{center}
\caption{Dynamic effective anisotropy: A time-harmonic source located
  inside a square array of pitch 2 consisting of 196 circular
  constrained holes of radius 0.4 leads to a wave pattern resembling a
  Saint-Andrews' cross at frequency $\Omega=1.966$ (a) FEM (b) HFH,
  and a Saint-George's cross at frequency $\Omega=2.75$ (c) FEM (d)
  HFH.}
\label{fig:cross}
\end{figure} 

\begin{figure}
  \begin{center}
    \includegraphics[scale=0.8]{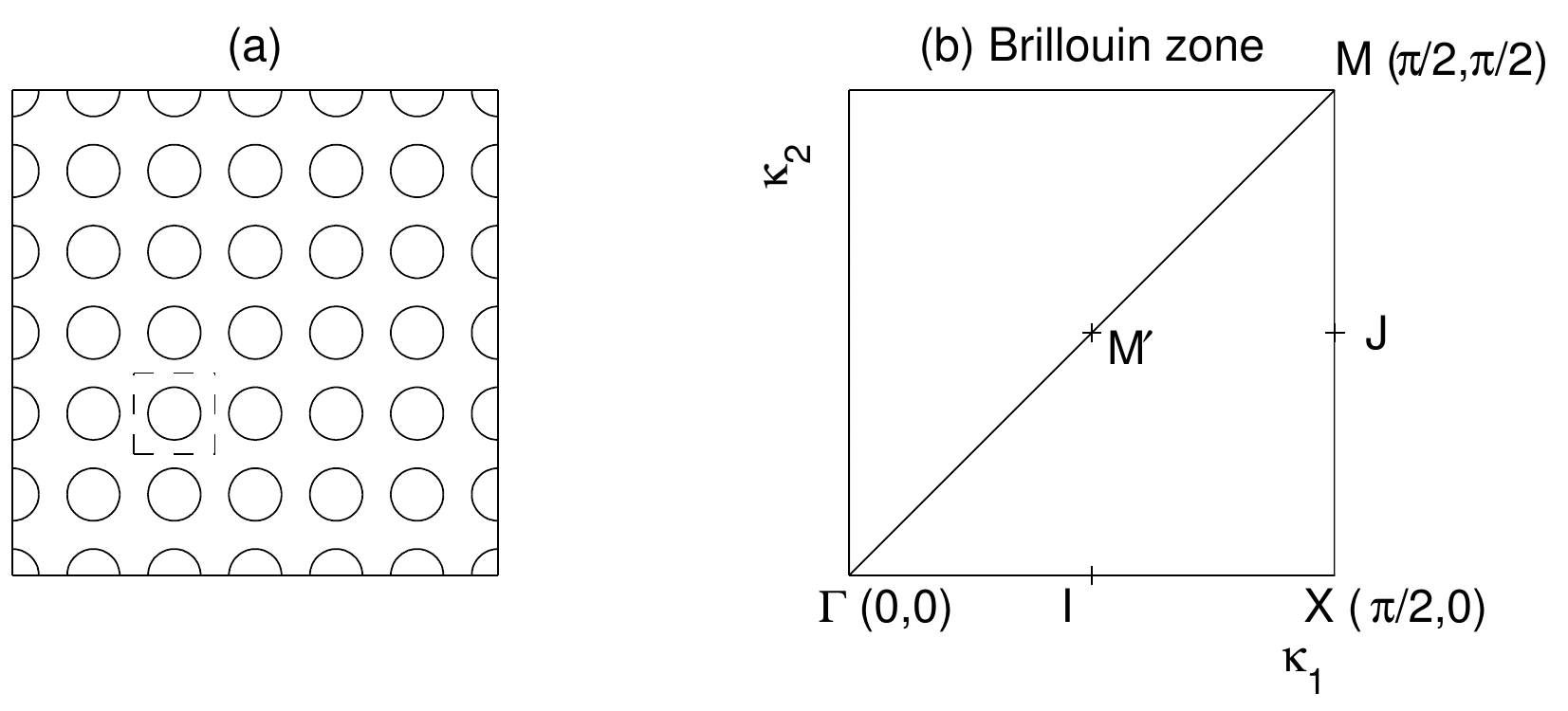} 
   \end{center}
\caption{Panel (a) shows a square array of circular cylinders, with the
  elementary cell as the dashed box. In (b) the irreducible Brillouin
  zone is shown together with lettering necessary for our discussion
  of folding in section \ref{sec:folds}. 
}
\label{fig:brillouin}
\end{figure}

 We show in Fig. \ref{fig:homo} 
what HFH does in practice: It focuses its attention on the physics
within a supercell of sidelength $L$, which is the long- scale, and
further captures the fine features of field's oscillations inside an
elementary cell of sidelength $2l$ much smaller than $L$
\cite{craster10a}. The wavelength need not be large compared to $l$ in
order to perform the asymptotic analysis, as the small parameter
$\epsilon=l/L$ only requires the supercell to be much larger than
its constituent elementary cells; this contrasts with classical
homogenization that assumes the cells are much smaller than the
wavelength. In this way, one replaces a periodic structure by a
homogenized one, on the long-scale, at any frequency and  classical homogenization is just a 
particular case of HFH \cite{antonakakis13a}.  An 
illustration of the HFH theory is in Fig. \ref{fig:cross}; the left
panels are from full finite element simulations of an array of
cylindrical holes (infinite
conducting boundary conditions of the holes which are for transverse
magnetic (TM) waves in
electromagnetism, or clamped shear horizontal waves in elasticity) excited at the array centre by a line source and the right
panels replace the array with its effective HFH medium which reproduces
the large scale behaviour. 
For the frequencies chosen, two very
distinct cross-shaped propagations, a Saint Andrews' cross (or
saltire) Fig. \ref{fig:cross}(a),(b), and a Saint George's cross
Fig. \ref{fig:cross} (c),(d) are achieved; these distinctive shapes
are created by excitations at precise frequencies that are 
identified from the dispersion curves, \cite{craster12b}, this strong frequency dependent
anisotropy is a recurrent feature of PCs. One sees that the HFH neatly reproduces
in Fig. \ref{fig:cross}(b,d) the fine features of the full finite
element simulations. Interpretation is provided by using the Brillouin zone
of Fig. \ref{fig:brillouin} and dispersion curves shown in Fig. \ref{fig:2D_Dirichlet}. The square array of holes behaves as two
dramatically contrasting effective media at frequency $\Omega=1.966$,
see Fig. \ref{fig:cross}(a,b), which stands on the edge of the
zero-frequency stop band that coincides with the lower edge of the
second stop band; and at frequency $\Omega=2.75$ which stands on the
upper edge of the second stop band, wherein the dispersion curve is
nearly flat is the $MX$ direction; both these frequencies are 
well beyond the range of applicability of classical homogenization.


Our aim is to generate HFH for arrays of holes with Dirichlet
conditions, so this is for TM waves in electromagnetism, and then compare through full
numerical simulations of HFH and finite elements to show how the
 behaviour of the numerous interesting   optical effects emerge naturally through the
 coefficients of our effective medium.  This is intertwined with a
 knowledge and understanding of the wave structure for a perfect
 medium where the dispersion relations connecting the phase shift
 across an elementary cell with the frequency is key. In section
 \ref{sec:general} we set up the mathematical framework leading to the
 effective medium equation (\ref{eq:T2D}), this is in a
 non-dimensional setting so for clarity in section \ref{sec:eff_media} the main
 results are summarized in a dimensional setting. There are degenerate
 cases that are of substantial interest and connect with Dirac-like
 dispersion \cite{chan12a} and these lead to a different effective
 equation also outlined in section \ref{sec:eff_media}. 

A first step in verifying the HFH theory is by asymptotically
reproducing the dispersion curves for perfect infinite arrays, and we
use both circular and square holes for illustration, as shown in section
\ref{sec:Disp}. Local to the edges of the Brillouin zone there are
standing wave frequencies and the local asymptotic behaviour is given
from the effective medium equation. Importantly, one can then predict
{\it apriori}, from the signs of the HFH coefficients with the effective
equation, how the bulk medium will behave. As a further illustration,
holes that degenerate to a point are treated (section \ref{sec:zero}) as there is then an exact
and simple solution for the dispersion relation, furthermore in this
case degeneracies occur. Finally, in section \ref{sec:examples} we
compare HFH with full numerical simulations showing the full range of
features available and how HFH represents them. Some concluding
remarks are drawn together in section \ref{sec:conclude}. 

\section{General theory}
\label{sec:general}
A two dimensional structure composed of a doubly periodic,
i.e. periodic in both $x$ and $y$ directions, 
square array of cells  with, not necessarily circular, identical holes
inside them is considered (see Fig. \ref{fig:brillouin}). We are primarily concerned with
electromagnetism where Maxwell's equations separate into transverse
electric (TE) and TM
polarizations and the natural boundary conditions on the holes are
then Neumann and
Dirichlet respectively; the asymptotics for the former are considered in
\cite{antonakakis13a} 
and applications in metamaterials are
illustrated with split ring resonators \cite{pendry99a}. In the current
article our emphasis is rather different, and is
upon TM polarization for which the Dirichlet conditions induce
different effects, such as zero-frequency stop bands, and require a modified theory. 

Assuming an $\exp(-i\omega
t)$ time dependence that is considered understood, and henceforth
suppressed, the governing equation is,
\begin{equation}
\nabla_{\bx}\cdot [\hat{a}(\bxi)\nabla_{\bx}u(\bx)]+\omega^2\hat{\rho}(\bxi)u(\bx)=0
\label{eq:2d}
\end{equation}
on $-\infty<x_1,x_2<\infty$ where $\omega$ is the angular frequency. In the periodic setting, each cell is
identical and the material is characterized by two
periodic functions on the short-scale, in $\bxi\equiv (x_1/l,x_2/l)$, namely
$\hat{a}(\bxi)$ and $\hat{\rho}(\bxi)$. In the context of photonics, these are
the inverse of permeability ($\mu^{-1}$), and permittivity ($\varepsilon$),
respectively, which are related to the wavespeed of light in vacuum $c$
via the refractive index $n$ as follows: $\varepsilon\mu=n^2 c^{-2}$.
The unknown $u$ physically is
the longitudinal component $E_z$ of the out-of-plane electric field ${\bf E}=(0,0,E_z)$,
bearing in mind that the transverse magnetic field ${\bf H}=(H_x,H_y,0)$ is retrieved
from Maxwell's equation ${\bf H}=i\omega^{-1} \mu^{-1} \nabla\times{\bf E}$.
In the context of phononics, $\hat{a}(\bxi)$ and $\hat{\rho}(\bxi)$ would
stand respectively for the shear modulus and the density
of an isotropic elastic medium, and the unknown $u$
would correspond to the out-of-plane displacement
(amplitude of SH shear waves). One can also
easily draw analogies with acoustic pressure
waves in a fluid, or linear surface water
waves, which explains the activity related to finding 
correspondences between models of electromagnetic \cite{sar_rpp05}
and acoustic \cite{craster12a} metamaterials. 
The length of the direct
lattice base vectors are taken equal and to be
$2l$, and define a
short lengthscale. The overall dimension of the structure is of a much longer length-scale, $L$; the
ratio of scales, $\epsilon\equiv l/L\ll 1$ then provides a small parameter
for use later in the asymptotic scheme. 

A non-dimensionalization of the physical
functions by setting, $\hat{a}\equiv\hat{a}_{0}a(\bxi)$ and
$\hat{\rho}\equiv\hat{\rho}_{0}\rho(\bxi)$ is convenient. Equation
(\ref{eq:2d}) then becomes
\begin{equation}
l^2\nabla_{\bx}\cdot
[a(\bxi)\nabla_{\bx}u(\bx)]+\Omega^2\rho(\bxi)u(\bx)=0 \quad {\rm with} \quad \Omega = \frac{\omega l}{\hat{c}_{0}}
\label{eq:2dnormal}
\end{equation}
 as the non-dimensional frequency and  $\hat{c}_0=\sqrt{\hat{a}_0/\hat{\rho}_0}$. The two scales, $l,
L$, then motivate two new coordinates namely ${\bf
  X}=\bx/L$, and $\bxi=\bx/l$ that are treated as independent, and placing this change of coordinates into (\ref{eq:2dnormal}) we obtain,
\begin{eqnarray}
&\nabla_{\bxi}\cdot[a(\bxi)\nabla_{\bxi}u({\bf
      X},\bxi)]+\Omega^2\rho(\bxi) u ({\bf X},\bxi)
  \nonumber\\
  &\qquad +\epsilon[2a(\bxi)\nabla_\bxi
    +\nabla_{\bxi}a(\bxi)]\cdot\nabla_{{\bf X}}u({\bf X},\bxi)
  +\epsilon^2 a(\bxi)\nabla^2_{\bf X}u({\bf X},\bxi)=0.
\label{eq:two-scales2D}
\end{eqnarray}
For wave propagation through a perfect lattice there exist standing waves at specific
eigenfrequencies, $\Omega_0$, and their eigenmodes satisfy periodic, or
anti-periodic (out-of-phase) boundary conditions on the edges of the
cell: We obtain asymptotic solutions based upon these standing waves. 


A key point, from this separation of scales, is that the boundary
conditions on the short-scale are known. 
For instance, periodic conditions in $\bxi$ on the edges of the cell are
\begin{equation}
u|_{\xi_{i}=1}=u|_{\xi_{i}=-1} \quad {\rm and} \quad u_{,\xi_i}|_{\xi_{i}=1}=u_{,\xi_i}|_{\xi_{i}=-1}
\label{eq:periodicBC}
\end{equation}
where $u_{,\xi_i}$ denotes partial differentiation with respect to $\xi_i$.
An almost identical analysis holds near the anti-periodic standing
wave frequencies: We illustrate the periodic case for
definiteness. On the long-scale we set no boundary conditions, and
indeed the methodology we present can be used even when the problem is
not perfectly periodic \cite{craster11a}. 

The following ansatz is taken 
\begin{equation}
u({\bf X},\bxi)=u_0({\bf X},\bxi)+\epsilon u_1({\bf X},\bxi)+\epsilon^2 u_2({\bf X},\bxi)+\ldots , \quad \Omega^2=\Omega_0^2+\epsilon \Omega_1^2+\epsilon^2 \Omega_2^2+\ldots:
\label{eq:expansion2D}
\end{equation}
Since $u({\bf X},\bxi)$ is periodic in $\bxi$ so are the $u_i({\bf
  X},\bxi)$'s.
 This leads to a hierarchy of equations in increasing powers of $\epsilon$:
 The leading order followed by the first order, second order and so
on. The first three equations in ascending order yield
\begin{equation}
(au_{0,\xi_i})_{,\xi_i} + \Omega_0^2\rho u_0=0
\label{eq:leadingOrder}
\end{equation}
\begin{equation}
(au_{1,\xi_i})_{,\xi_i} + \Omega_0^2\rho u_1=
-(2au_{0,\xi_i}+a_{,\xi_i}u_0)_{,X_i}
-\Omega_1^2 \rho u_0
\label{eq:firstOrder}
\end{equation}
\begin{eqnarray}
  & (au_{2,\xi_i})_{,\xi_i} + \Omega_0^2\rho u_2
  \nonumber\\
  &\qquad =-au_{0,X_iX_i}
   -(2au_{1,\xi_i}+ a_{,\xi_i}u_1)_{,X_i}-\Omega_1^2\rho u_1
   -\Omega_2^2\rho u_0
.
  \label{eq:secondOrder}
\end{eqnarray}
We start by expressing a solution for the leading order equation. For
a specific eigenvalue $\Omega_0$ there is a corresponding eigenmode
$U_0(\bxi;\Omega_0)$,  periodic in $\bxi$, and the leading order solution is 
\begin{equation}
 u_0({\bf X},\bxi)=f_0({\bf X}) U_0(\bxi;\Omega_0).
\label{eq:LeadingSolution}
\end{equation} 
The function $f_0({\bf X})$
 is determined later on, indeed the whole point is to deduce
a long-scale continuum partial differential equation for $f_0$
entirely upon the long-scale. We begin with a
treatment of isolated eigenvalues, however repeated eigenvalues also 
arise, and require modifications, and will be dealt with
in sections \ref{sec:RepEigen} and \ref{sec:RepEigenQuad}: These are
relevant to Dirac-like cones \cite{chan12a}. 

Dirichlet conditions on the inside boundary of the cell, $\partial
S_2$, where $S$ denotes the surface of the cell in the $\bxi$ coordinates, yield
\begin{equation}
u({\bf X},\bxi)|_{\partial S_2}=0 \iff u_i({\bf X},\bxi)|_{\partial S_2}=0, \quad i\in\ \mathbb{N}
\label{eq:dirichletConditions}
\end{equation}
 and are set in the short-scale $\bxi$ so,  
for $i=0$, $U_0(\bxi;\Omega_0)|_{\partial S_2}=0$.

To deduce the long-scale PDE for $f_0({\bf X})$ we follow
\cite{craster10a}, making changes where necessary due to the change in
boundary conditions. Equation (\ref{eq:firstOrder}) is multiplied by
$U_0$ and integrated over the cell's surface to obtain
\begin{eqnarray}
   & \dint_S \left( U_0(a u_{1,\xi_i})_{,\xi_i} +\Omega_0^2\rho
      U_0 u_1\right) dS \nonumber\\
  &\qquad\qquad =-f_{0,X_i} \dint_S(a
    U_0^2)_{,\xi_i} dS -f_0\Omega_1^2 \dint_S\rho U_0^2 dS .
  \label{eq:intermed1}
\end{eqnarray}
The first integral of the right hand side is converted to a path integral along $\partial S$ using a corollary of the divergence theorem. The periodic conditions on $\partial S_1$ and the homogeneous Dirichlet conditions of $U_0$ on $\partial S_2$ make it vanish. We continue by subtracting the integral over the cell of the product of equation (\ref{eq:leadingOrder}) with $u_1/f_0$ to obtain
\begin{equation}
  \dint_S (U_0(a u_{1,\xi_i})_{,\xi_i}
      - u_1(a U_{0,\xi_i})_{,\xi_i}) dS
    = -f_0\Omega_1^2 \dint_S \rho(\bxi)U_0^2 dS.
\label{eq:intermed2}
\end{equation}
Using Green's theorem equation (\ref{eq:intermed2}) becomes
\begin{equation}
 \int_{\partial S}  a(\bxi)\left(U_0\frac{\partial u_1}{\partial
     n}-u_1\frac{\partial U_0}{\partial n}\right) ds=-f_0\Omega_1^2 \dint_S \rho(\bxi)U_0^2 dS
\label{eq:intermed3}
\end{equation}
 where $\partial S=\partial S_1 +\partial S_2$. 
Due to periodic (or anti-periodic) boundary conditions of $u$ and its first derivatives
with respect to $\xi_i$ on $\partial S_1$, and due to the homogeneous
Dirichlet boundary conditions of the $u_i$'s on $\partial S_2$, the
left hand side of equation (\ref{eq:intermed3}) vanishes. It follows
that $\Omega_1=0$, which is an important deduction as it implies that
the asymptotic behaviour of dispersion curves near isolated
eigenfrequencies is at least quadratic. 
We then solve for $u_1({\bf X},\bxi)$ and obtain 
\begin{equation}
u_1({\bf X},\bxi)=f_1({\bf X}) U_0(\bxi;\Omega_0)
    +\nabla_{\bf X} f_0({\bf X})\cdot {\bf U_1}(\bxi).
\label{eq:u1Solution}
\end{equation}
 The first term is simply a homogeneous solution that is irrelevant, and the vector
 function ${\bf U}_1$ is a particular solution. 
Quite often exact solutions for $u_1$ or ${\bf U}_1$ are not possible
to obtain so one must use numerical solutions and this is described in
section \ref{sec:examples}. 
We now move to the second order equation to compute $f_0({\bf X})$ and $\Omega_2^2$.
Similarly to the first order equation, we multiply equation
(\ref{eq:secondOrder}) by $U_0$ then subtract the product of equation
(\ref{eq:firstOrder}) with $u_2/f_0$, finally we integrate over the
cell's surface to obtain the following effective equation for  $f_0$ 
\begin{equation}
  T_{ij} f_{0,X_iX_j}+\Omega_2^2 f_0=0,\quad {\rm with}\quad
  T_{ij}=\frac{t_{ij}}{\dint_S \rho U_0^2 dS}\quad {\rm for} \quad i,j=1,2,
\label{eq:T2D}
\end{equation}
 which is the main result of this article. In particular, the
 coefficients $T_{ij}$ encode the anisotropy at a specific
 frequency and the $t_{ij}$'s are integrals over the small-scale cell
\begin{equation}
t_{ii}=\dint_SaU_0^2dS+2\dint_SaU_{1_i,\xi_i}U_0dS+\dint_S a_{,\xi_i}U_{1_i}U_0dS,
\label{eq:t11}
\end{equation}
\begin{equation}
t_{ij}=2\dint_SaU_{1_j,\xi_i}U_0dS+\dint_S a_{,\xi_i}U_{1_j}U_0dS \quad {\rm for} \quad i\neq j,
\label{eq:tij}
\end{equation}
 where $U_{1_i}$ is the ith component of vector function ${\bf U}_1$.
Note that there is no summation over repeated indexes for $t_{ii}$. 
 The PDE for $f_0$, equation (\ref{eq:T2D}), is crucial as the local
 microstructure is completely encapsulated within the tensor $T_{ij}$;
 these are, for a specific structure at an $\Omega_0$, just numerical
 values as illustrated in table \ref{tab:first_six}. Notably the tensor can
 have negative values, or components, and it  allows one to
 interpret and, even more importantly, predict changes in behaviour or
 when specific effects occur. The structure of the tensor depends upon
 the boundary conditions of the holes, the results here, for instance,
 are different from those of the Neumann case
 \cite{antonakakis13a}. Another key point is that numerically the
 short scale is no longer present and the PDE (\ref{eq:T2D}) is simple and quick to solve
 numerically, or even by hand. 


One primary aim of the present article is to deduce formulae for the
local behaviour of dispersion curves near standing wave
frequencies as these shed light upon the physical effects observed. For this one returns to the perfect lattice and then 
 Floquet Bloch boundary conditions on the elementary cell lead
 immediately to $f_0({\bf X})=\exp(i\kappa_j X_j/\epsilon)$. In this notation
 $\kappa_j=K_j-d_j$ and $d_j=0,\pi/2,-\pi/2$ 
depending on the location in the Brillouin zone. Equation (\ref{eq:T2D}) gives
\begin{equation}
\Omega\sim\Omega_0+ \frac{T_{ij}}{2\Omega_0}\kappa_i \kappa_j
\label{eq:asymptoticExpansion}
\end{equation} 
 and these locally quadratic dispersion curves are completely
 described by $\Omega_0$ and the tensor $T_{ij}$.

\subsection{Classical singularly perturbed zero-frequency limit}
\label{sec:longwave}
It is natural, at this point, to contrast with usual homogenization
theories. As is well-known \cite{nicorovici95a,mcphedran97a} one
cannot homogenize the TM polarized case because the conventional
approach only works at low frequencies in a quasi-static limit.
Some authors attribute this to a singular perturbation, or
non-commuting limit, problem whereby letting first
the Bloch wavenumber and then the frequency tend
to zero, or doing it the way around, leads to a different
result \cite{poulton01a}.
The approach used in this article overcomes this by releasing the theory
from the low-frequency constraint. For TE polarization
\cite{antonakakis13a} one can
explicitly connect the theories and show that the low frequency theory
is a sub-set of the high frequency approach. The TM case differs
fundamentally as
there is a zero-frequency stop-band. 

We now prove that the usual homogenization cannot give the asymptotic
behaviour even at low-frequency. Setting 
$\Omega^2=\epsilon^2\Omega_2^2+\ldots$, that is, going to low frequency, and assuming that $u_0({\bf
  X},\bm\xi)=f({\bf X})U_0({\bm\xi})$ is non-zero one arrives at a
contradiction: At leading order  $U_0(\bm\xi)$ satisfies
\(
 \nabla^2 U_0=0
\)  
  where for brevity we have set $a=1$, 
  complemented by $U_0=0$ on the hole and periodic (anti-periodic)
 boundary condition on the edge of the cell. From the uniqueness
 properties of the Laplacian $U_0\equiv 0$ is the
 unique solution and hence usual homogenization cannot find the
 asymptotics at low frequency in this Dirichlet case. 


\subsection{Repeated eigenvalues: linear asymptotics}
\label{sec:RepEigen}
Repeated eigenvalues are commonplace in specific examples and, as we
shall see, are particularly relevant to Dirac-like cones
\cite{chan12a} that have
practical significance. If we assume repeated eigenvalues of multiplicity $p$ the general solution for the leading order problem becomes,
\begin{equation}
u_0=f_0^{(l)}({\bf X})U_0^{(l)}(\bxi;\Omega_0)
\label{eq:uzeroRep}
\end{equation}
 where we sum over the repeated superscripts $(l)$. 
Proceeding as before, we multiply equation (\ref{eq:firstOrder}) by $U_0^{(m)}$, subtract $u_1((aU^{(m)}_{0,\xi_i})_{\xi_i}+\Omega_0^2\rho U_0^{(m)})$ then integrate over the cell to obtain,
\begin{equation}
f_{0,X_i}^{(l)}\dint_SU_0^{(m)}(2aU_{0,\xi_i}^{(l)}+a_{\xi_i}U_0^{(l)})dS+\Omega_1^2 f_0^{(l)}\dint_S\rho U_0^{(l)}U_0^{(m)}dS=0.
\label{eq:lam1}
\end{equation}
 There is now a system of coupled partial differential equations for
 the $f_0^{(l)}$ and, provided $\Omega_1\neq 0$, the leading order
 behaviour of the dispersion curves near the $\Omega_0$ is now linear
 (these then form Dirac-like cones).  These coupled partial differential
 equations on the long-scale now replace (\ref{eq:T2D}) near these frequencies.

For the perfect lattice and Bloch wave problem, we set $f_0^{(l)}={\hat
  f}_0^{(l)}\exp(i\kappa_jX_j/\epsilon)$ and 
 obtain the following equations,
\begin{equation}
\left(i\frac{\kappa_j}{\epsilon}{\bf A}_{jml}+\Omega_1^2{\bf
    B}_{ml}\right){\hat f}_0^{(l)}=0,  \quad {\rm for} \quad m=1,2,...,p
\label{eq:PreSystem}
\end{equation}
where,
\begin{equation}
{\bf A}_{jml}=\dint_SU_0^{(m)}(2aU_{0,\xi_j}^{(l)}+a_{,\xi_j}U_0^{(l)})dS
\label{eq:Amatrix}
\end{equation}
and
\begin{equation}
{\bf B}_{ml}=\dint_S\rho U_0^{(l)}U_0^{(m)}dS.
\label{eq:Bmatrix}
\end{equation}
The system of equations (\ref{eq:PreSystem}) is written simply as,
\begin{equation}
{\bf C}{\hat {\bf F}}_0=0,
\label{eq:System}
\end{equation}
with ${\bf C}_{ll}=\Omega_1^2{\bf B}_{ll}$ and ${\bf
  C}_{ml}=i\kappa_j{\bf A}_{jml}/\epsilon$ for $l\neq m$. One must then solve
for $\Omega_1^2=\pm\sqrt{\alpha_{ij}\kappa_i\kappa_j}/\epsilon$ when the determinant of ${\bf C}$ vanishes and the asymptotic relation is,
\begin{equation}
\Omega\sim\Omega_0\pm\frac{1}{2\Omega_0}\sqrt{\alpha_{ij}\kappa_i\kappa_j}.
\label{eq:asymptoticExpansionLinear}
\end{equation}
If $\Omega_1$ is zero, one must go to the next order and a slightly
different analysis ensues.

\begin{table}
\centering
\begin{tabular}{ccc}\hline
$\alpha_{11}$ & $\alpha_{22}$ & $ \Omega_0$ \\
$2\pi^2$ & $2\pi^2$ & $\pi $\\
$4\pi^2$ & $4\pi^2$ & $\sqrt{2}\pi $\\
\hline
\end{tabular}
\caption{The coefficients $\alpha_{ii}$, necessary in equation (\ref{eq:asymptoticExpansionLinear}) for the two Dirac-like cones at point $\Gamma$ of the Brillouin zone for a doubly periodic array of constrained points.}
\label{tab:r_0_01_alphas}
\end{table}

\subsection{Repeated eigenvalues: Quadratic asymptotics}
\label{sec:RepEigenQuad}
Assuming that $\Omega_1$ is zero, $u_1=f_{0,X_k}^{(l)}U_{1_k}^{(l)}$ (we again
sum over all repeated $(l)$ superscripts) and 
 advance to second order using
(\ref{eq:secondOrder}). Taking the difference between
the product of equation (\ref{eq:secondOrder}) with $U_0^{(m)}$ and
$u_2((aU_{0,\xi_i})_{,\xi_i} + \Omega_0^2\rho U_0)$ and then
integrating 
over the elementary cell gives
 \begin{eqnarray}
&f_{0,X_iX_i}^{(l)}\dint_SaU_0^{(m)}U_0^{(l)}dS+
f_{0,X_kX_j}^{(l)}\dint_SU_0^{(m)}(2aU_{1_k,\xi_j}^{(l)}+a_{,\xi_j}U_{1_k}^{(l)})dS
\nonumber\\
&\quad +\Omega_2^2 f_0^{(l)}\dint_S\rho U_0^{(m)}U_0^{(l)}dS=0, \quad
{\rm for} \quad m=1,2,...,p
\label{eq:preLam2}
\end{eqnarray}
 as a system of coupled PDEs. 
The above equation is presented more neatly as 
\begin{equation}
f_{0,X_iX_i}^{(l)}{\bf A}_{ml}+f_{0,X_kX_j}^{(l)}{\bf
  D}_{kjml}+\Omega_2^2 f_0^{(l)}{\bf B}_{ml}=0, \quad {\rm for} \quad m=1,2,...,p.
\label{eq:Lam2}
\end{equation}
For the Bloch wave setting, using $f_0^{(l)}({\bf X})={\hat f}_0^{(l)}\exp(i\kappa_jX_j/\epsilon)$ we obtain the following system,
\begin{equation}
\left(-\frac{\kappa_i\kappa_i}{\epsilon^2}{\bf A}_{ml}-\frac{\kappa_k\kappa_j}{\epsilon^2}{\bf
  D}_{kjml}+\Omega_2^2{\bf B}_{ml}\right){\hat f}_0^{(l)}=0, \quad {\rm for} \quad m=1,2,...,p
\label{eq:sys3}
\end{equation}
 and this determines the asymptotic dispersion curves.

\section{Effective dispersive media}
\label{sec:eff_media}
The major goal of HFH is to represent the periodic medium of finite
extent using an effective homogeneous medium and the main result of this
article is in achieving this, and for clarity this is summarized in
this section back
in the dimensional setting. 

Transforming equation (\ref{eq:T2D}) back to the original $x_i=X_iL$ coordinates and using the solution of $\Omega_2$ we obtain an effective medium equation for ${\hat f}_0$ that is,
\begin{equation}
T_{ij}{\hat f}_{0,x_ix_j}({\bf x})+\frac{\Omega^2-\Omega_0^2}{l^2}{\hat f}_0({\bf x})=0.
\label{eq:fEff}
\end{equation}
The nature of equation (\ref{eq:fEff}) is not necessarily elliptic
since $T_{11}$ and $T_{22}$ can take different values and/or different
signs. In the illustrations herein the $T_{ij}=0$ for $i\neq j$. The hyperbolic behaviour of equation (\ref{eq:fEff}) yields asymptotic 
solutions that describe the endoscope effects where, if one of the
$T_{ii}$ coefficients is zero as often happens near the point $M$ of
the Brillouin zone, waves propagate only in one direction. Note that
when the cell has the adequate symmetries the dispersion relation near
point $G(0,\pi/2)$ is identical to that near point $X(\pi/2,0)$ which
explains the existence of two orthogonal directions of propagation
instead of only one. 

\begin{figure}
  \begin{center}
    \includegraphics[scale=0.8]{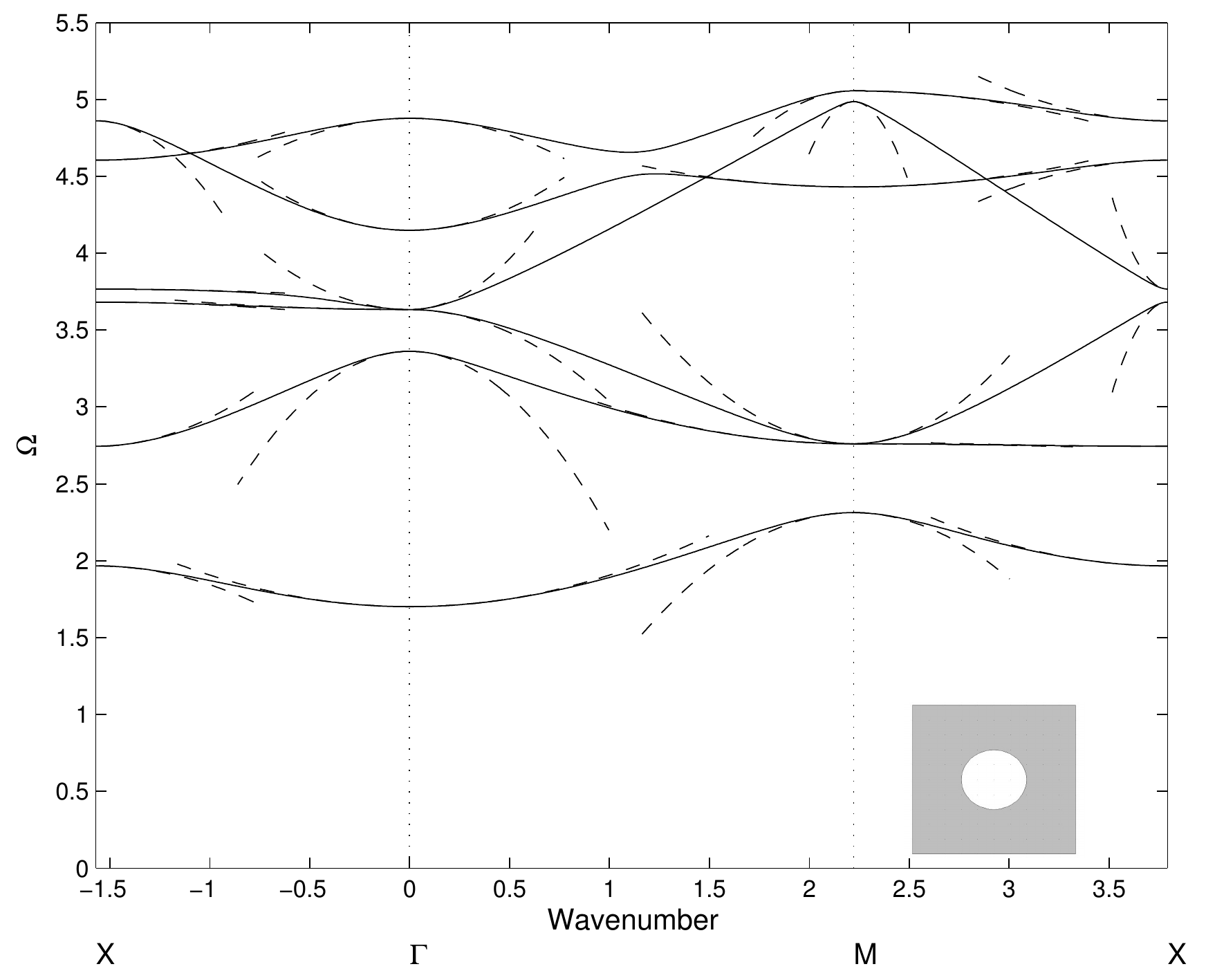}
   \end{center}
\caption{The dispersion diagram for a doubly periodic array of square cells with circular inclusions, of radius $0.4$, fixed at their inner boundaries shown for the irreducible Brillouin zone of
  Fig. \ref{fig:brillouin}. The dispersion curves are shown in solid lines and the asymptotic solutions from HFH are shown in dashed lines.}
\label{fig:2D_Dirichlet}
\end{figure}

\begin{table}
\centering
\begin{tabular}{ccc}\hline
$T_{11}$ & $T_{22}$ & $ \Omega_0$ \\
$0.698841854976085$ & $0.698841854976085$ & $1.700916617638699 $\\
$7.867675441589871$ &  $7.867675441589871$ & $3.361627184739501 $\\
$0.3230$ & $-8.998$ & $3.632730109763024 $  \\
$4.9681$ &  $14.2899$ & $3.632730129014608 $\\
$4.775527931399718$ &  $4.775527931399718$ & $4.148661549527329 $\\ 
$-4.279843054160115$ & $-4.279843054160115$ & $4.877003447953185 $\\
\hline
\end{tabular}
\caption{The first six standing wave frequencies for in-phase waves at
  $\Gamma$,
  cf. Fig. \ref{fig:2D_Dirichlet}, together with associated values for
  $T_{11}$ and $T_{22}$. Symmetry between $T_{11}$ and $T_{22}$ is breaking when the multiplicity of the eigenvalue is greater than two. Negative group velocity is demonstrated by the negative sign of both $T_{ij}$'s.}
\label{tab:first_six}
\end{table}

At Dirac-like cones the linear behaviour of the effective medium yields an
equation slightly different from (\ref{eq:fEff}). Regarding the zero
radius holes, presented in the following section \ref{sec:zero}, it consists of a system of three coupled PDE's that uncouple to yield one identical PDE for all $f_0^{(i)}$s that is of the form,
\begin{equation}
{\hat f}_{0,x_ix_i}({\bf x})+\beta\frac{(\Omega^2-\Omega_0^2)^2}{l^2}{\hat f}_0({\bf x})=0,
\label{eq:effDirac}
\end{equation}
where $\beta$ is a coefficient equivalent of the $T_{ij}$ but this
time is the same for all combinations of $i$ and $j$. The quadratic
mode that emerges in the middle of the linear ones follows equation
(\ref{eq:fEff}).

\section{Dispersion curves}
\label{sec:Disp}
We now generate dispersion curves for circular and square holes and
verify the HFH theory versus these numerically generated curves; this
is a good test of the approach as the dispersion curves contain
changes in curvature, coalescing branches and modes that
cross. These dispersion curves can then be used to interpret the
physical optics phenomena seen later. 
Most vividly the dispersion curves also show the zero frequency, and
other, stop bands.

\subsection{Circular inclusions in a square array}
\label{sec:DirichletInclusion}
The circular inclusions are arranged in a square array and each
elementary cell is a square of side $2$, we consider large holes
initially and then how the curves change as the radius
decreases. Ultimately we consider infinitesimal radii and perhaps
remarkably obtain explicit solutions. In almost all other cases one
has to use fully numerical techniques, or semi-analytical methods such
as multipoles \cite{zolla05a,movchan02a}.

\subsubsection{Large circular holes}
\label{sec:largeHoles}
\begin{figure}
  \begin{center}
    \includegraphics[scale=0.7]{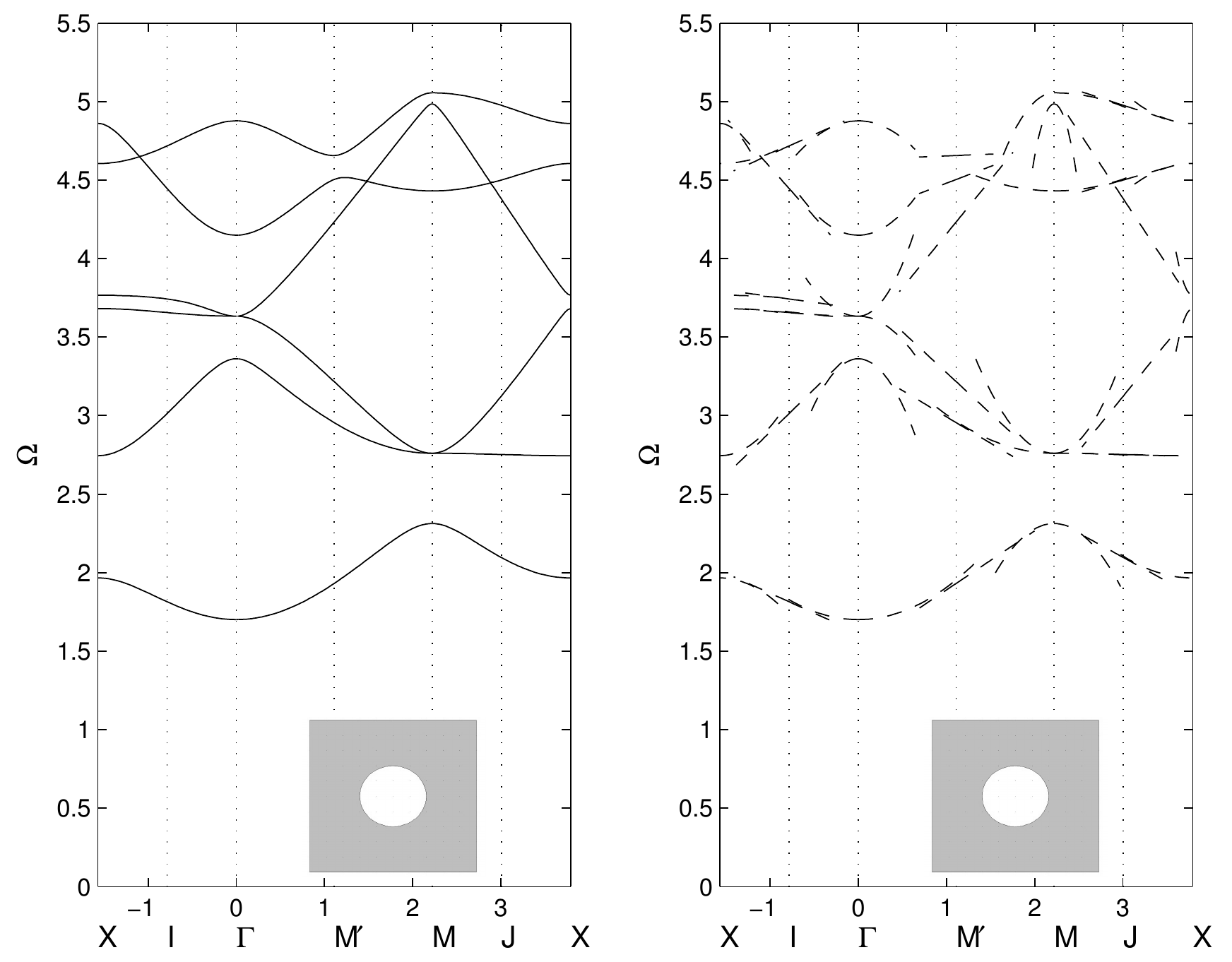} 
   \end{center}
\caption{The solid line dispersion curves are from FE computations (circle of radius
  $0.4$) and the dashed are from the HFH asymptotics. These curves
  become virtually indistinguishable when one takes larger and larger
  macro cells, as this amounts to perturbing away from an increasing
  number of standing waves on the diagrams, which is the essence of
  the folding technique. In the limit of an infinite macro cell one
  would perturb away from a dense set of points on the diagrams, and
  their reconstruction would become perfect \protect\cite{kirill-personal}. 
}
\label{fig:folds}
\end{figure}

The full dispersion curves are computed numerically
using COMSOL (finite element code \cite{comsol}) and then asymptotically with the HFH 
 using the equations developed in section
\ref{sec:general}. Fig. \ref{fig:2D_Dirichlet}
illustrates the typical dispersion curves versus the asymptotics for
wavenumbers, on the specified path of Fig. \ref{fig:brillouin}, for
the example of a
hole with radius $0.4$. There is, as expected, a zero-frequency stop-band and the lowest
branch is isolated with a full stop-band also lying above it. At some
standing wave frequencies two modes share the same frequency, for
instance the third and fourth modes at $\Gamma$; these
modes are asymptotically quadratic in the local wavenumber. Table \ref{tab:first_six} shows
the $T_{11}, T_{22}$ values for point $\Gamma$ of the Brillouin zone. Note how
$T_{11}=T_{22}$ for all single eigenfrequencies, but that symmetry
breaks for the double roots. Moreover the signs of the $T_{11}$ and
$T_{22}$ naturally inform one of the local curvature near $\Gamma$. Physically, this tells us the sign of group velocity of waves with small phase-shift across the unit cells, and thus whether or not they undergo backward propagation, which is one of the hallmarks of negative refraction.

\subsubsection{Folding technique for reconstructing the dispersion curves asymptotically}
\label{sec:folds}  

An apparent deficiency of the approach we present is that it requires
known standing wave frequencies, and associated eigenstates, at the
band edges  and that
the asymptotics may be poor at frequencies far from these standing
waves; until now HFH, applied to the dispersion
curves, has been limited to obtaining asymptotic solutions near the
standing wave frequencies \cite{craster10a}. This uses an 
elementary cell containing a single hole and the irreducible Brillouin zone associated with it,
however by using larger macro-cells containing four or more holes, and foldings of their 
resultant Brillouin zone, one can extract the asymptotics at other
positions in wavenumber space thereby considerably enhancing the asymptotically
coverage, see Fig. \ref{fig:folds}.  As an example, we take an elementary square cell of length
$2$ with circular holes of radius $0.4$, section 
\ref{sec:DirichletInclusion} and Fig. \ref{fig:2D_Dirichlet}.

To obtain asymptotic solutions at the wavenumber positions, $I(\pi/4,0)$, $J(\pi/2,\pi/4)$, $M'(\pi/4,\pi/4)$
of Fig. \ref{fig:folds}(b), bisecting the three segments $[\Gamma X]$, $[MX]$ and $[\Gamma M]$ of the Brillouin zone
 we can use the equations (\ref{eq:T2D}, \ref{eq:t11}, \ref{eq:tij}) again, section
 \ref{sec:RepEigen},  but now with different 
elementary cells.

 Let the Brillouin zone
for a square macro-cell composed of four holes be
$\Gamma'M'I$. At $\Gamma'(0,0)$ periodic (in-phase) conditions on the
macro cell boundaries yield all in-phase and out-of-phase modes of the
elementary cell containing a single hole. This elementary cell's $[\Gamma M]$ segment is
folded on point $M'$. At $M'$, where anti-periodic (out-of-phase)
conditions on the macro cell are applied, the dispersion curves
contain repeated roots, which are linear, and their slope is
equal to the asymptote of the dispersion curves of the elementary cell
setting at point $M'$. Proceeding in a similar manner one can fold the
Brillouin zone at will by considering macro-cells composed of
simple unfoldings of the elementary cell. Asymptotics can be obtained
for wavenumber positions of the Brillouin zone equal to all ratios of
the elementary cell's Brillouin zone wavenumber points. These ratios
 are equal to $1/n$ where $n$ is related to the number of
elementary cells of one hole contained in the macro cell in each
dimension. 

We now illustrate this by obtaining asymptotics at points $M'$, $I$ and
$J$ which respectively correspond to the following macro cell
settings, a square macro cell composed of four elementary cells with
anti-periodic boundary conditions, a rectangular macro cell composed
of two elementary cells in the $\xi_1$ direction with anti-periodic
boundary conditions on $\xi_1$ but periodic on $\xi_2$ and finally a
rectangular macro cell composed of two elementary cells in the $\xi_2$
direction with anti-periodic conditions in $\xi_1$ and $\xi_2$. The
asymptotics at points $M'$, $I$, $J$ as well as the usual points
$\Gamma$, $M$ and $X$ are illustrated in Fig. \ref{fig:folds}. 


\begin{figure}
  \begin{center}
    \includegraphics[scale=0.7]{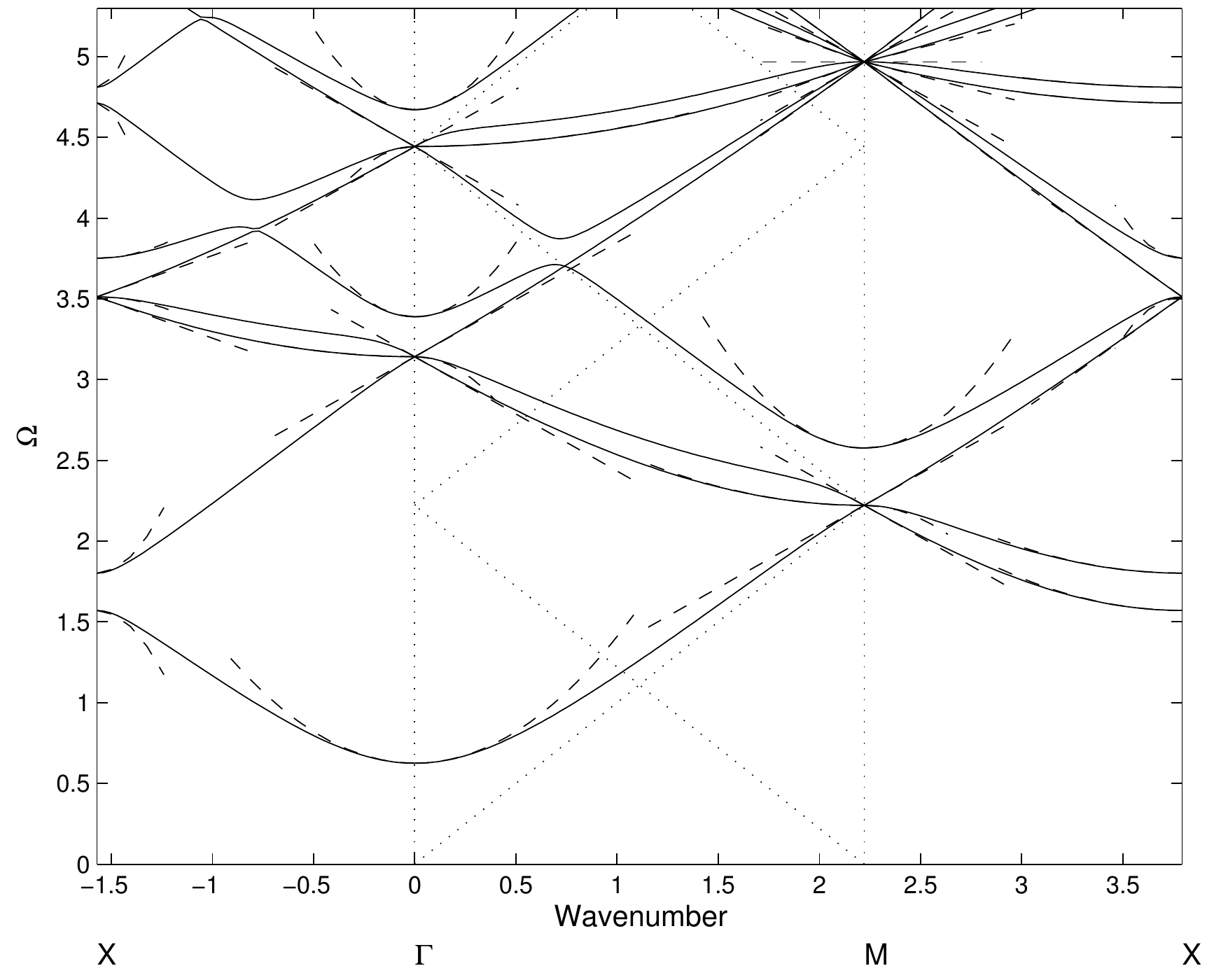}
   \end{center}
\caption{The dispersion curves for a doubly periodic array of
  constrained points: Solid lines are from exact dispersion relation
  (\ref{eq:Fourier}) and the dashed lines
  are HFH asymptotics. Additionally, we draw two sets of 
   dotted lines emerging from $\Gamma$ and $X$ which are folded light
   lines. 
}
\label{fig:zero}
\end{figure}

\begin{table}
\centering
\begin{tabular}{ccc}\hline
$T_{11}$ & $T_{22}$ & $ \Omega_0$ \\
$0.979719694474189$ & $0.979724590049377$ & $0.624563144241832 $\\
$1$ &  $-22.34$ & $\pi$\\
$12.439613406598808$ & $12.439765169617527$ & $3.388815608872719 $  \\
$-16.77$ &  $18.77$ & $\sqrt{2}\pi$\\
$18.856092144598160$ &  $18.855405067087230$ & $4.670779386112907 $\\
\hline
\end{tabular}
\caption{The first five frequencies for in-phase standing waves  and their respective $T_{ij}$ coefficients for a doubly periodic array of square cells with zero radius holes (see Fig. \ref{fig:zero}). Equal $T_{11}$ and $T_{22}$ coefficients are for the single multiplicity eigenmodes. The second and fourth set of $T_{ij}$ are for the quadratic modes of the Dirac-like cones.}
\label{tab:r_0_01_Tii}
\end{table}

\subsection{Circular inclusions of small radius}
\label{sec:r_0_01}

As the hole radius decreases, the dispersion curves gradually
transition toward those of the infinitesimal holes shown in Fig.
\ref{fig:zero}: The zero frequency stop-band is generic, and a feature
forced by the Dirichlet boundary condition, however, the lowest 
dispersion curve is no longer isolated and triple crossings created by
repeated roots occur.

For the larger hole radius, section \ref{sec:DirichletInclusion}, all
the modes are locally quadratic close to the standing wave
frequencies. As the wavenumber moves away from those specific
Brillouin zone points the behaviour of the dispersion curves becomes
locally linear. The formation of triple, and more, crossings as the radius
tends to zero, are created by the linear behaviour of the dispersion
curves far away from points $\Gamma$, $M$ and $X$ moving toward those
points to replace the quadratic behaviour of all the multiple modes,
except one, as seen in Fig. \ref{fig:zero}. Triple crossings are
illustrated at the second and fourth standing wave frequencies at
point $\Gamma$ and the first standing wave frequencies at point $M$,
where the third frequency at point $M$ yields a hep-tuple crossing.
Reassuringly, the asymptotic theory captures these multiple modes
whether the crossings are double, triple, or more as is illustrated in
Figs. \ref{fig:2D_Dirichlet}, \ref{fig:zero} where the circular
holes are respectively of radius $0$ and $0.4$. For the multiple
crossing points with multiplicity higher than two, both linear and
quadratic dispersion curves arise. One can proceed, as in section
\ref{sec:RepEigen}, to obtain the different $\Omega_1$ coefficients
and one of them is zero; proceeding to higher order and apply
equations (\ref{eq:preLam2}) to (\ref{eq:sys3}) gives the quadratic
behaviour of the middle mode emerging from the triple crossings.
 Again the $T_{ii}$ contain the critical information about the local
 behaviour, some typical values along $\Gamma$ are given in table
 \ref{tab:r_0_01_Tii}, for the quadratic curves; they are naturally
 identical for isolated modes but describe the inherent anisotropy for
 the other cases. One can then immediately see whether the material
 will behave with directional preferences at specific frequencies. 

Dispersion diagrams of zero radius and for $0.1$ radius (not shown)
geometries are almost indistinguishable by eye, but apparent triple
crossings in the latter are actually a double crossing together with a
single mode that has almost the same standing wave frequency. The
separation of the two almost touching standing wave frequencies
depends on the size of the inclusion and tends to zero as the
inclusion shrinks down to a point: Fig. \ref{fig:r_0_01_zoom}
illustrates a close-up of a triple point comparing inclusions of
radius $0$ and $0.1$. This is actually important for the Dirac-like cones
and associated effects.

\subsection{Zero radius, Fourier series and Dirac-like cones}
\label{sec:zero}
A doubly periodic array of constrained points (circles of zero radius)
is an attractive limit and has an immediate solution in terms of Fourier series
as
\begin{equation}
 u(\xi_1,\xi_2)=e^{i{\bm \kappa}\cdot{\bm\xi}}\sum_{n_1=-\infty}^\infty\sum_{n_2=-\infty}^\infty \frac{e^{-i\pi
   {\bf N}\cdot{\bm\xi}}}{(\kappa_1-\pi n_1)^2+(\kappa_2-\pi
 n_2)^2-\Omega^2}
\label{eq:ubloch}
\end{equation}
 with ${\bm\kappa}=(\kappa_1,\kappa_2)$ and ${\bf N}=(n_1,n_2)$ and 
 $\Omega$ and $\bm\kappa$ are related via the dispersion
 relation 
\begin{equation}
D({\bm\kappa},\Omega)=\sum_{n_1=-\infty}^\infty\sum_{n_2=-\infty}^\infty
\frac{1}{(\kappa_1-\pi
   n_1)^2+(\kappa_2-\pi n_2)^2-\Omega^2}=0.
\label{eq:Fourier}
\end{equation}
This is valid provided the double sum does not have a singularity, 
which  would occur whenever 
\begin{equation}
\Omega=\sqrt{(\kappa_1-\pi n_1)^2+(\kappa_2-\pi n_2)^2}.
\label{eq:singular}
\end{equation}
 These relations from the singularities also play a role in the
 dispersion picture; they correspond to the Bloch states of a perfect
 medium without any constraints (a homogeneous isotropic medium). However, in some circumstances they
 also, by serendipity, exactly satisfy  the constraint $u=0$ at the
 centre of each cell and therefore, in those cases, are simultaneously 
 dispersion curves; this is the origin of the degeneracy seen in the
 multiple crossings (occurring at Dirac-like points) of the dispersion curves of Fig. \ref{fig:zero}. 
It is also clear from the dispersion relation that 
 $\Omega=0$ is not a solution at $\bm\kappa={\bf 0}$ and hence that
 there is a zero-frequency stopband and conventional long-wave
 homogenization is doomed to failure cf. section \ref{sec:longwave}.

 A set of dispersion curves are shown in Fig. \ref{fig:zero}: The
 Bloch states for the perfect medium free of defects lead to light
lines folded at the edges of the Brillouin zone and those emerging
from $\Gamma$ are pertinent to the current discussion. These light
lines intersect at the multiple crossings, which
 are called generalized Dirac-like points: Generalized Dirac-like points occur
 for a given set of frequencies at all three edges of the Brillouin
 zone, namely for points $\Gamma$, $M$ and $X$. It is useful to find
 criteria for their multiplicity: At $\Gamma$ the generalized Dirac-like
 points occur for $\Omega^2=\pi^2m$ such that $m=n_1^2+n_2^2$ with $m$
 integer; the multiplicity depends on the ways in which $n_1$ and
 $n_2$ can form $m$. From
 elementary number theory $m$ can be equal to the sum of two squares
 in more than one way, i.e., $m=50=5^2+5^2=7^2+1^2$ or the sum of two squares can also be a
 perfect square, i.e., $m=25=5^2=4^2+3^2$. Equation
 $u_{,x_ix_i}+\pi^2(n_1^2+n_2^2)u=0$ together with the appropriate
 boundary conditions describes the generalized Dirac-like points and an
 independent set of solutions is formed from different possibilities
 of $\sin(\pi(n_ix+ n_jy))$, $\sin(\pi(n_ix-n_jy))$,
 $\cos(\pi(n_ix+n_jy))$-$\cos(\pi(n_px-n_ky))$ where $i,j,p,k \in
 \{1,2\}$, $i\neq j$ and $p\neq k$. The multiplicity can be equal to
 $3$, $7$, $11$ or more depending on $m$.  For example
 $\Omega^2=\pi^2$ with $n_i=1$, $n_j=0$ has multiplicity $3$ where
 $\Omega^2=5\pi^2$ has multiplicity $7$ and $\Omega^2=50\pi^2$ has
 multiplicity $11$. This is true for solutions at $\Gamma$ and $M$ but
 at $X$ one can obtain a singularity that accepts only one
 eigensolution. Due to the non-symmetry of the general sum at $X$ that
 renders the denominator of the Fourier series singular,
 $m=(1/2-n_1)^2+n_2^2$, it is possible to obtain solutions of
 multiplicity one as in the case of $\Omega_0=\pi/2$ where the only
 eigensolution respecting the boundary conditions is $\sin(\pi x/2)$.

Given the exact solution one can, in these special cases, generate the
asymptotics by hand. For the asymptotics we construct the coefficients
using $U_0$ from (\ref{eq:ubloch}) and deduce  ${\bf U}_1=(U_{11},U_{12})$ as
\begin{equation}
 U_{1k}=2ie^{i{\bm \kappa}\cdot{\bm\xi}}
\sum_{n_1=-\infty}^\infty\sum_{n_2=-\infty}^\infty
\frac{(\kappa_k-\pi n_k) e^{-i\pi
   {\bf N}\cdot{\bm\xi}}}{[(\kappa_1-\pi n_1)^2+(\kappa_2-\pi
 n_2)^2-\Omega^2]^2}
\end{equation}
 and thus one can extract the $T_{ij}$ in (\ref{eq:T2D}) by doing the
 integrals by hand, the off-diagonal terms are
 zero, and $T_{11}$ and $T_{22}$ are
\begin{equation}
   T_{11}=1-4\frac{\sum_{n_1=-\infty}^\infty\sum_{n_2=-\infty}^\infty
\frac{(\kappa_1-\pi n_1)^2}{[(\kappa_1-\pi n_1)^2+(\kappa_2-\pi n_2)^2-\Omega_0^2]^3}}
{\sum_{n_1=-\infty}^\infty\sum_{n_2=-\infty}^\infty
\frac{1}{[(\kappa_1-\pi n_1)^2+(\kappa_2-\pi n_2)^2-\Omega_0^2]^2}},\quad\\
\end{equation}
 and an identical equation for $T_{22}$ but with $1$ and $2$
 interchanged, 
 where $(\kappa_1,\kappa_2)$ are $(0,0)$, $(\pi/2,0)$ and $(\pi/2,\pi/2)$ at the
 edges of the Brillouin zone at the respective points of $\Gamma$, $X$ and $M$. These asymptotics only apply at the isolated,
 non-repeating roots, and are shown in Fig. \ref{fig:zero}.

\begin{figure}
  \centering
    \includegraphics[scale=0.3]{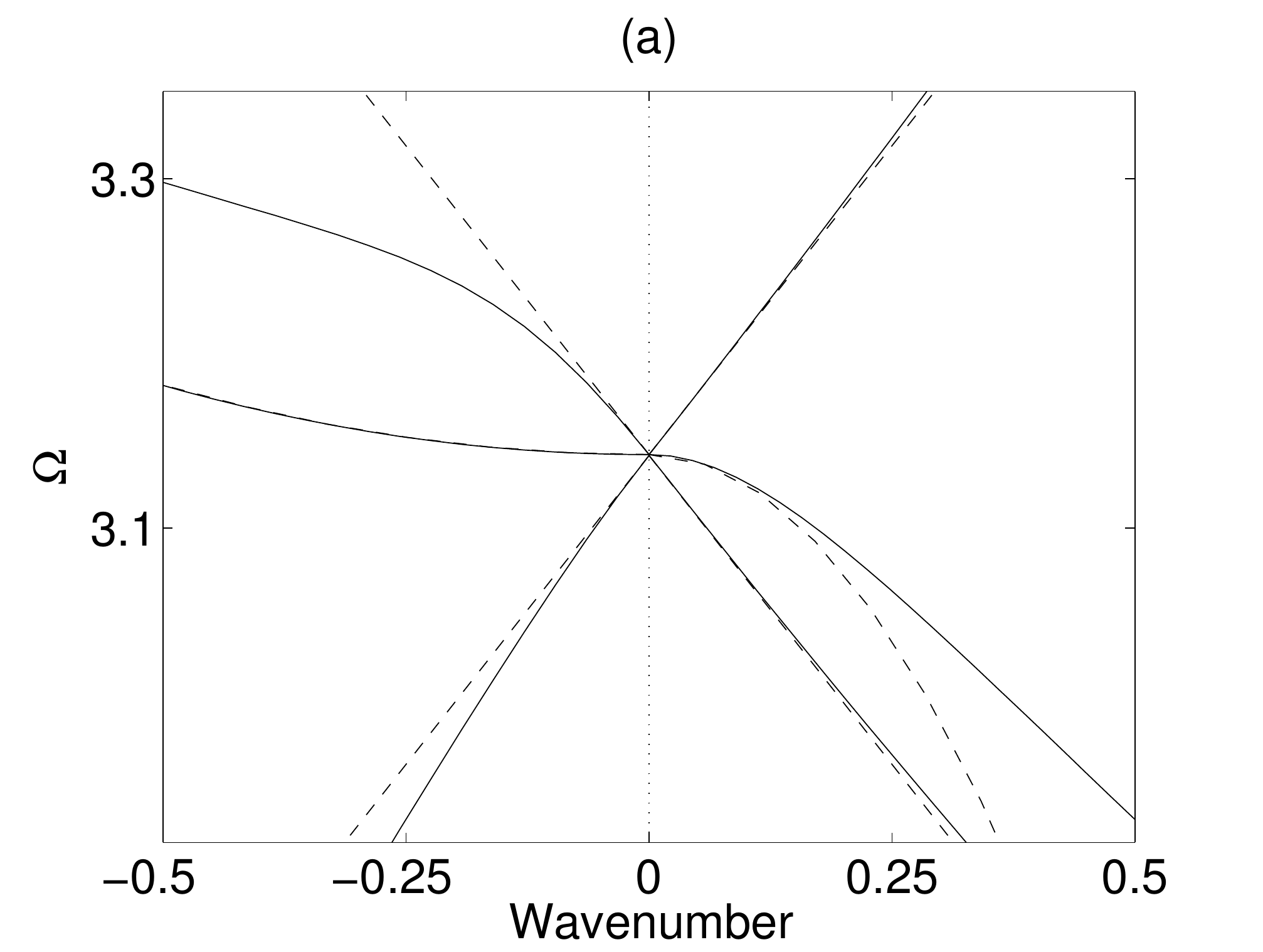} 
    \includegraphics[scale=0.3]{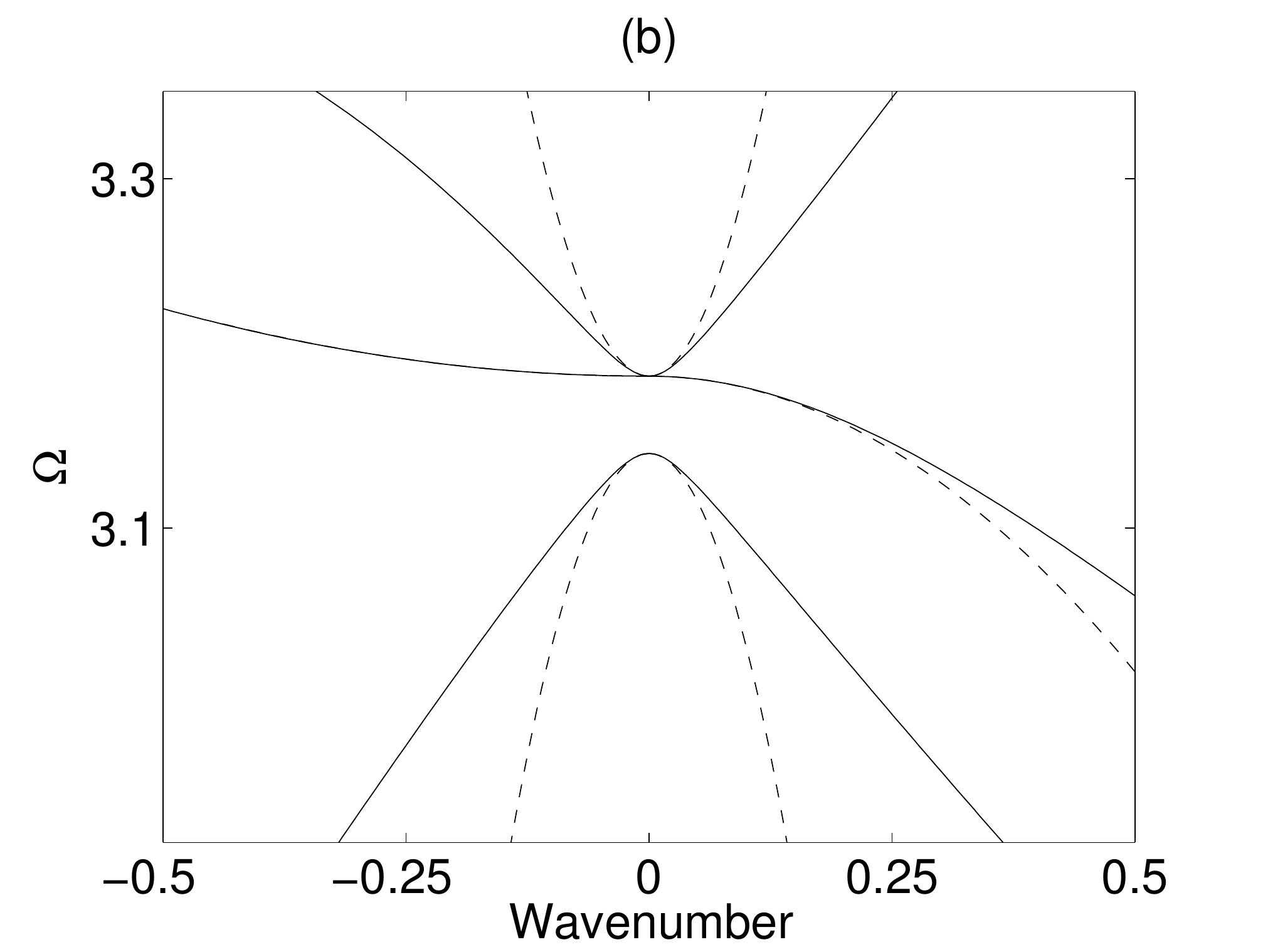}
\caption{Panel (a) shows a close up of the Dirac-like point and its asymptotics from figure \ref{fig:zero}. The values for the asymptotic coefficients of equation (\ref{eq:asymptoticExpansionLinear}) are $\alpha_{11}=\alpha_{22}=2\pi^2$ and for equation (\ref{eq:asymptoticExpansion}) they are $T_{11}=1$ and $T_{22}=-22.34$. Panel (b) shows the same region for a slightly larger hole radius of $0.1$ where one can see the merging of the double mode with the mode near $\pi$ to form a triplet as the radius decreases to zero. HFH captures both cases as is seen from the asymptotic curves. As the radius decreases the top and bottom quadratics become steeper and ultimately reach a linear behaviour.
}
\label{fig:r_0_01_zoom}
\end{figure}

The linear cases at the triple crossings are easily extracted from
section \ref{sec:RepEigen}. For instance, for the repeated case at $\Omega_0=\pi$ there are
three linearly independent solutions with 
\begin{equation} u_0=f^{(1)}({\bf X})\sin\pi \xi_1 +f^{(2)}({\bf X})\sin\pi \xi_2 
+f^{(3)}({\bf X})[\cos\pi \xi_1 -\cos\pi \xi_2]
\end{equation}
 Following through the methodology one arrives at
$2\Omega_1^4=(2\pi)^2(\kappa_1^2+\kappa_2^2)$ which gives the two linear
asymptotics, and similarly at the other points. Table
\ref{tab:r_0_01_alphas} summarizes the coefficients for the first two
Dirac-like cones at point $\Gamma$. The asymptotics are shown in Fig. \ref{fig:r_0_01_zoom}
 showing that the linear behaviour is perfectly captured. 

\begin{figure}
\begin{center}
\includegraphics[scale=0.8]{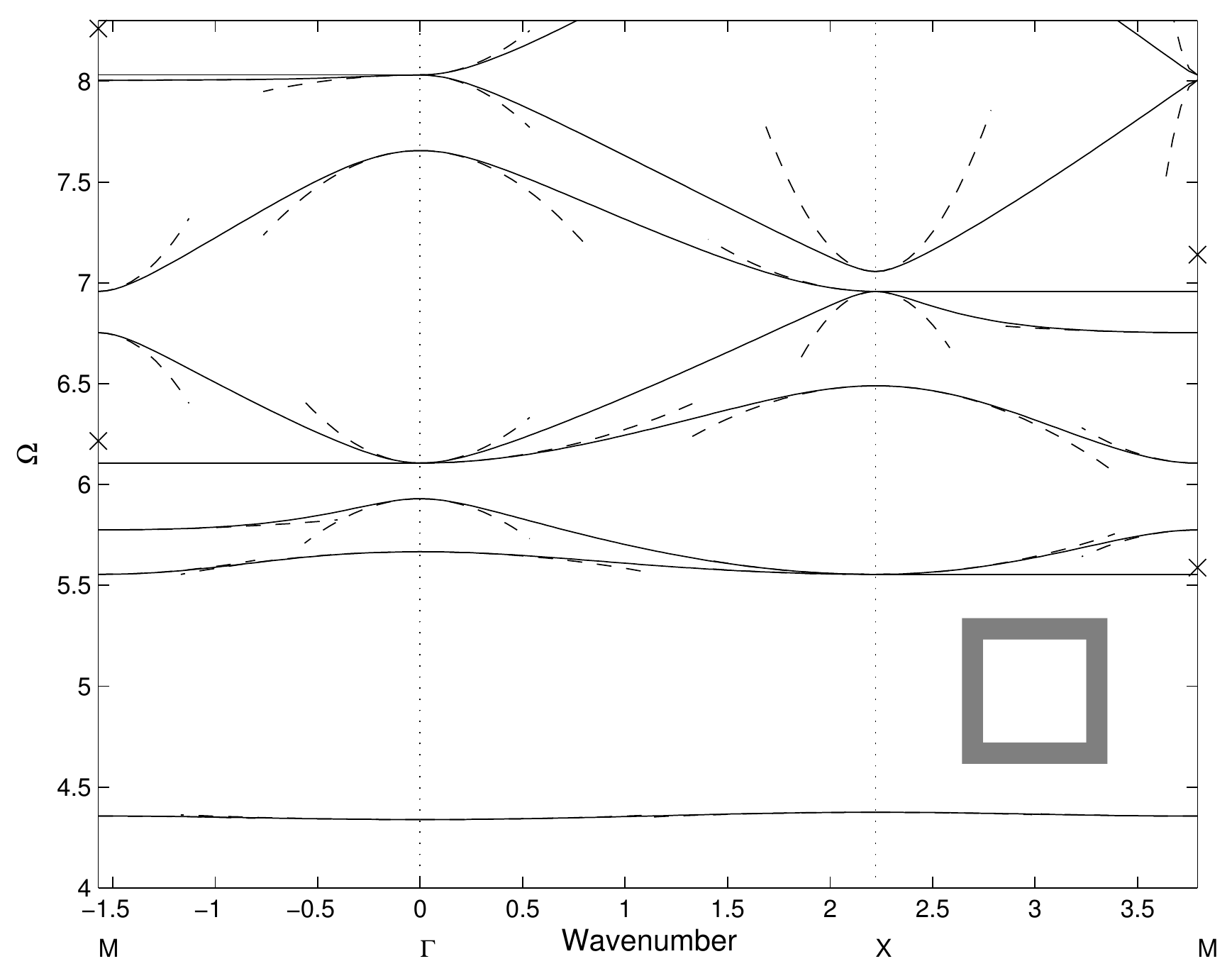}
\end{center}
\caption{The dispersion curves for square inclusions, of side $\sqrt 2$, from numerical
  simulation (solid) and from HFH theory (dashed). The HFH and numerics  for the
  lowest two curves are virtually indistinguishable. The crosses on the frequency axis are frequencies predicted
   by solving Helmholtz's equation in a waveguide consisting of a rectangle with Neumann data on one side and Dirichlet data on the other sides, see also Figure. 10. Their values in increasing order are $5.59$, $6.22$, $7.14$ and $8.26$.}
\label{tilt0}
\end{figure}

\begin{figure}
\begin{center}
\includegraphics[scale=0.8]{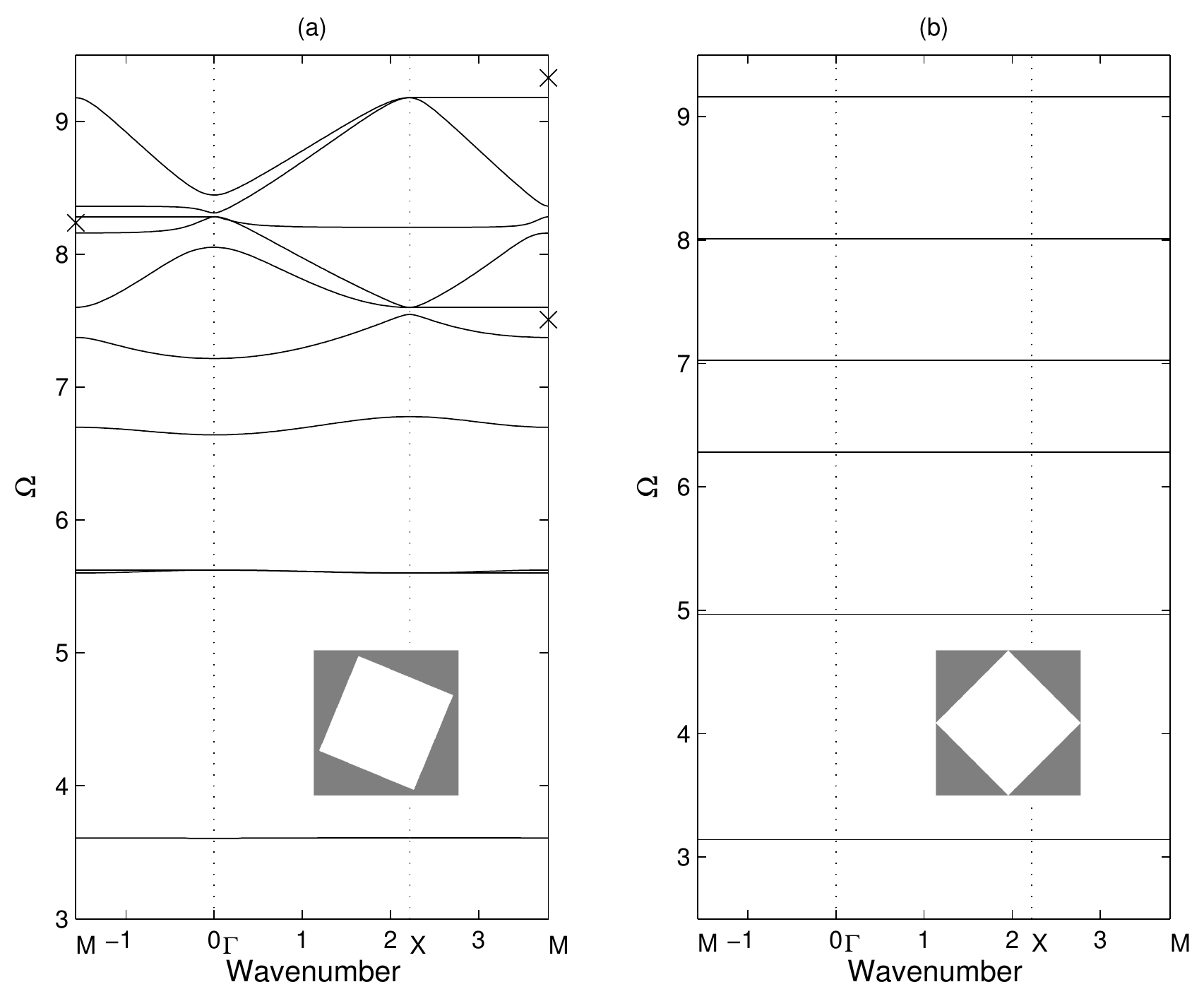}
\end{center}
\caption{The dispersion curves for a square of side $\sqrt{2}$ from
  numerical simulation for (a) a square rotated by $\pi/8$ and (b) by
  $\pi/4$. The crosses on the frequency axis in (a) are frequencies predicted
   by solving Helmholtz's equation in a waveguide. Their values are $7.51$, $8.24$ and $9.33$.}
\label{tiltpi4pi8}
\end{figure}

\begin{figure}
\begin{center}
\includegraphics[scale=0.6]{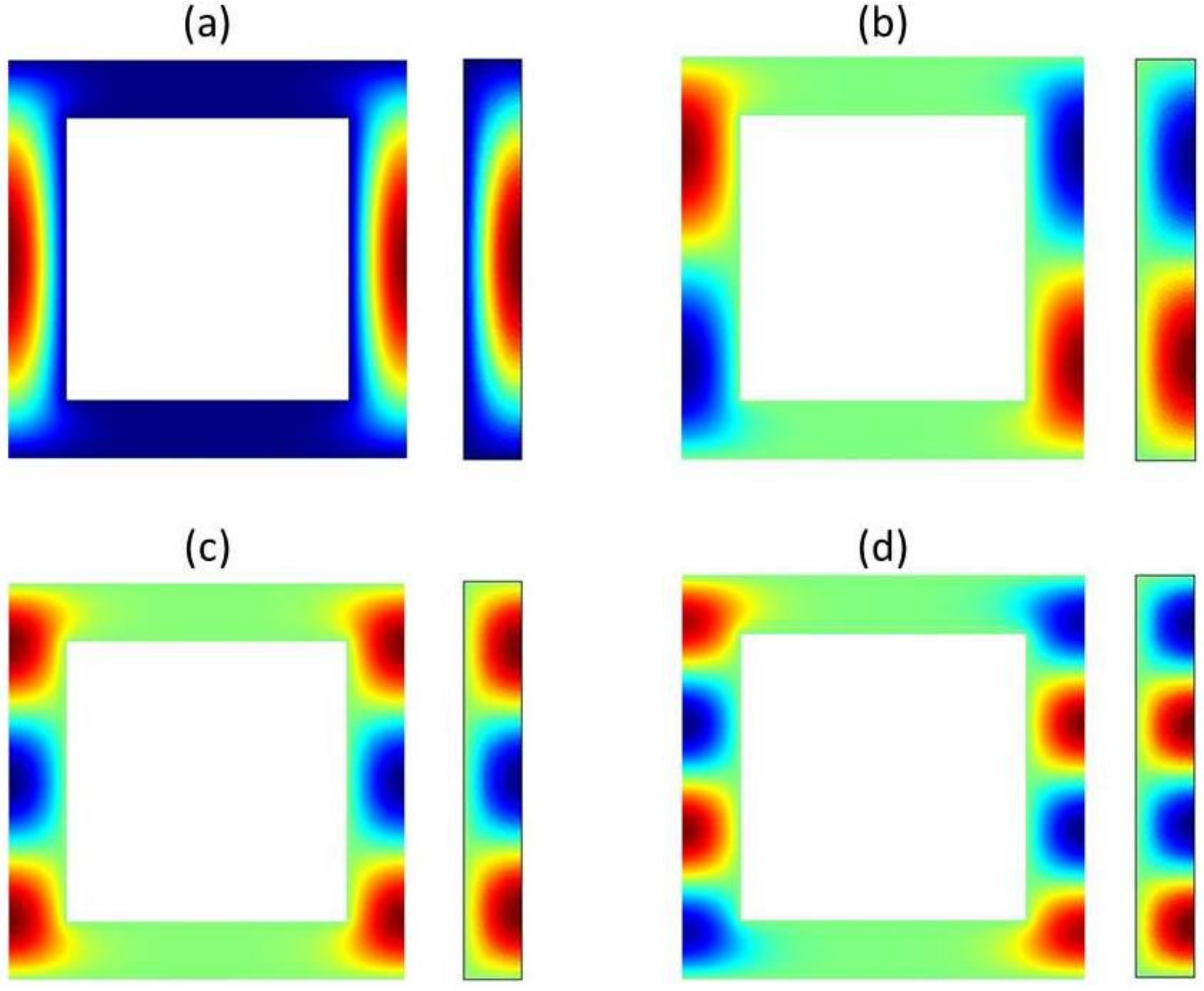}
\end{center}
\caption{The eigenstates of the first four flat bands at the
  respective frequencies of $5.56$, $6.11$, $6.96$ and $8.03$. The
  left of every panel shows FEM computations of the eigenstate, as
  noted in the text a waveguide model can be developed to approximate
  the field and eigenvalue and these waveguide counterparts are shown
  alongside.
}
\label{fig:StandingWaves}
\end{figure}

\subsection{Square inclusions}
\label{sec:square}
Our methodology is not limited to circular inclusions and is easily
applied to any shape, we briefly consider square inclusions taking a
square of side $\sqrt{2}$ and see in Fig. \ref{tilt0} that the dispersion
curves are not unlike those of the large cylinder in
Fig. \ref{fig:2D_Dirichlet}: an isolated lowest mode separating a zero
frequency stop-band from  a wide stop-band, yet another stop band
arises near $\Omega=6$. The asymptotics from HFH again reproduce the
behaviour near the edges of the Brillouin zone and can be used to
construct effective HFH equations. The HFH results in
Fig. \ref{fig:2D_Dirichlet} are almost completely indistinguishable
all along the lowest two branches and are highly accurate near the
edges of the Brillouin zone for the other branches. 
Interestingly, the orientation of
the square matters strongly and
rotating it progressively flattens the lowest (lower frequency) dispersion curves until
ultimately, for a rotation of $\pi/4$, they become completely flat
leading to resonant states: The rotation gradually isolates each piece
of material from its neighbours leaving resonant blocks that vibrate
with the eigenfrequency of Dirichlet squares of side $\sqrt 2$
\(
 \Omega^2=(m\pi/\sqrt 2)^2+(n\pi/ \sqrt 2)^2
\) 
 with $m, n$ non-zero integers. 
\subsubsection{Standing wave modes}
\label{sec:StandingWaves}
It is noticable in Figs. \ref{tilt0} and \ref{tiltpi4pi8}a that
completely flat modes exist which represent standing waves for a whole
range of Bloch wavenumbers. The eigenfunctions, for the non-tilted
case, 
for the first four standing wave
frequencies, 
are shown in Fig. \ref{fig:StandingWaves}; it is clear that the field
is concentrated along parallel and opposite sides and this suggests a
simple equivalent waveguide model can be constructed. 
 Taking a rectangle with a Neumann condition on one side and Dirichlet
conditions on the other three sides where  the Neumann condition on
one side is necessary to cover the appropriate symmetry. The length and width of the
rectangle depends on the tilt of the square hole. For the non-tilted
case the length is taken to be the length of a cell and the width
as the half width of the cell minus the inner square. The resulting
frequency estimate $\Omega_E$ reads, 
\begin{equation}
\Omega_E=\pi\left(\left(\frac{2n-1}{l-w}\right)^2+\left(\frac{m}{l}\right)^2\right)^{\frac{1}{2}}
\quad\text{for}\quad n,m \in \mathbb N^*
\label{eq:FreqEst}
\end{equation}
where $l$ and $w/2$ are respectively the effective length and width of the waveguide. For the non-tilted square hole $l_0=2$ and $w_0=2-\sqrt 2$ where for a tilt of $\pi/8$ they reduce to the values of $l_{\pi/8}=1.6051$ and $w_{\pi/8}=0.4335$. 
Equation (\ref{eq:FreqEst}) then leads to the
'x' crosses placed on Figs. \ref{tilt0} and \ref{tiltpi4pi8}a which are good
predictions of the standing wave frequencies, and the approximate
eigensolutions are shown in Fig. \ref{fig:StandingWaves}. 
The precision of the estimates 
 degenerates as the frequency  increases as the field near the 
 ends of the imaginary rectangle gradually leaks into the undisturbed ligaments.

\section{Lensing, guiding and cloaking effects}
\label{sec:examples}
We now connect the HFH theory with the exciting phenomena that are
topical in photonics.

\begin{figure}
\begin{center}
\includegraphics[scale=0.2]{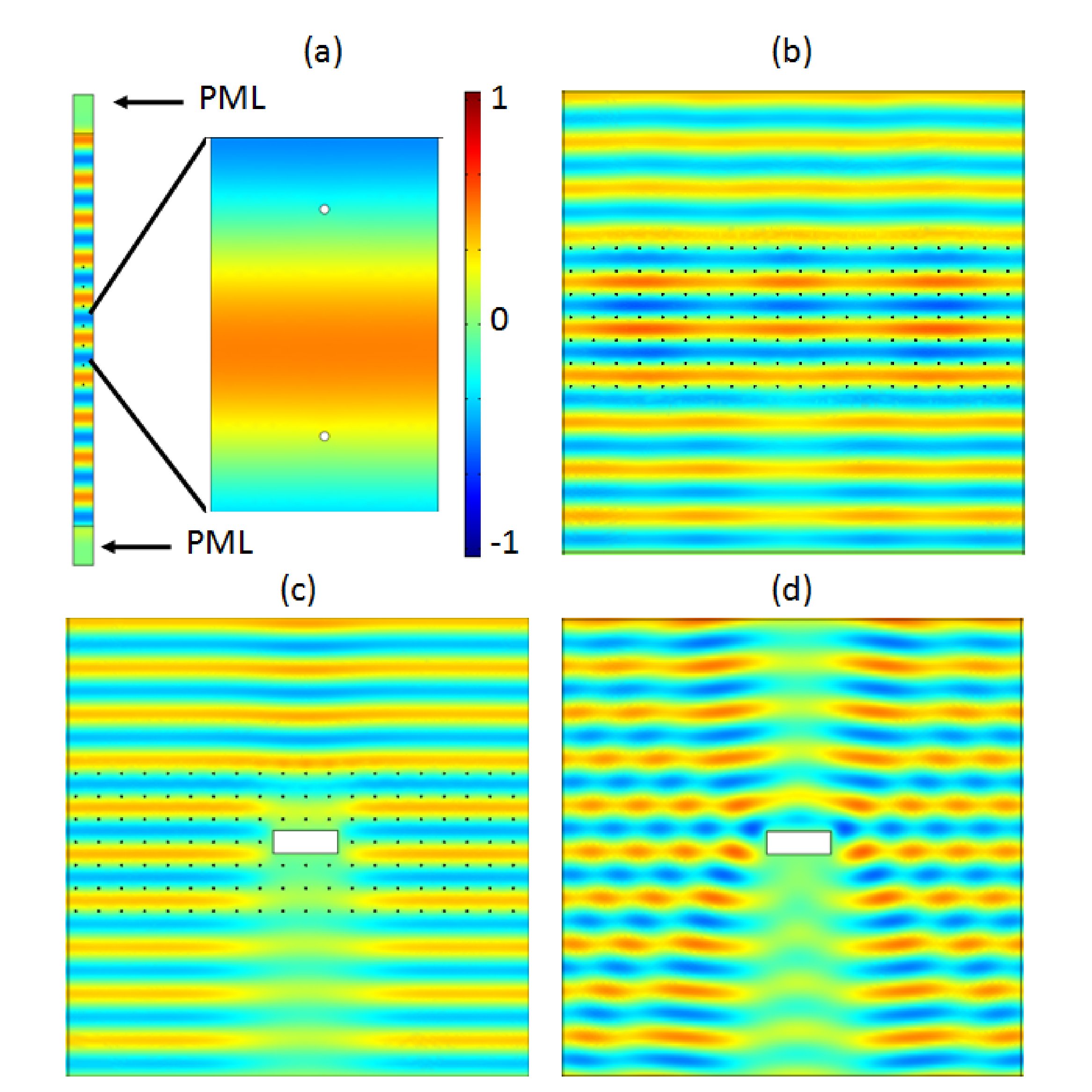}
\end{center}
\caption{Dirac-like cone versus cloaking: (a) Perfect transmission through a column
of seven constrained evenly spaced inclusions of radius $0.01$ (array pitch
$2$) with periodic conditions on either sides and Perfectly Matched Layers
(PML) on top and bottom, and a plane wave incident from above at frequency
$\Omega_0=2.215$; (b) Nearly perfect transmission for a plane wave
at frequency $\Omega_0$ incident from above on a finite array
of $140$ constrained inclusions with PML on either sides of
the computational domain; (c) Same as in panel (b) when a clamped
rectangular obstacle is placed inside the array; (d) Plane wave
incident from above on the clamped obstacle at frequency
$\Omega_0$ for comparison; The slightly
reduced amplitude in forward scattering in (c), but
the lack of backward scattering, is a hallmark
of cloaking. 
}
\label{figseb2}
\end{figure}

\begin{figure}
  \centering
    \includegraphics[scale=0.25]{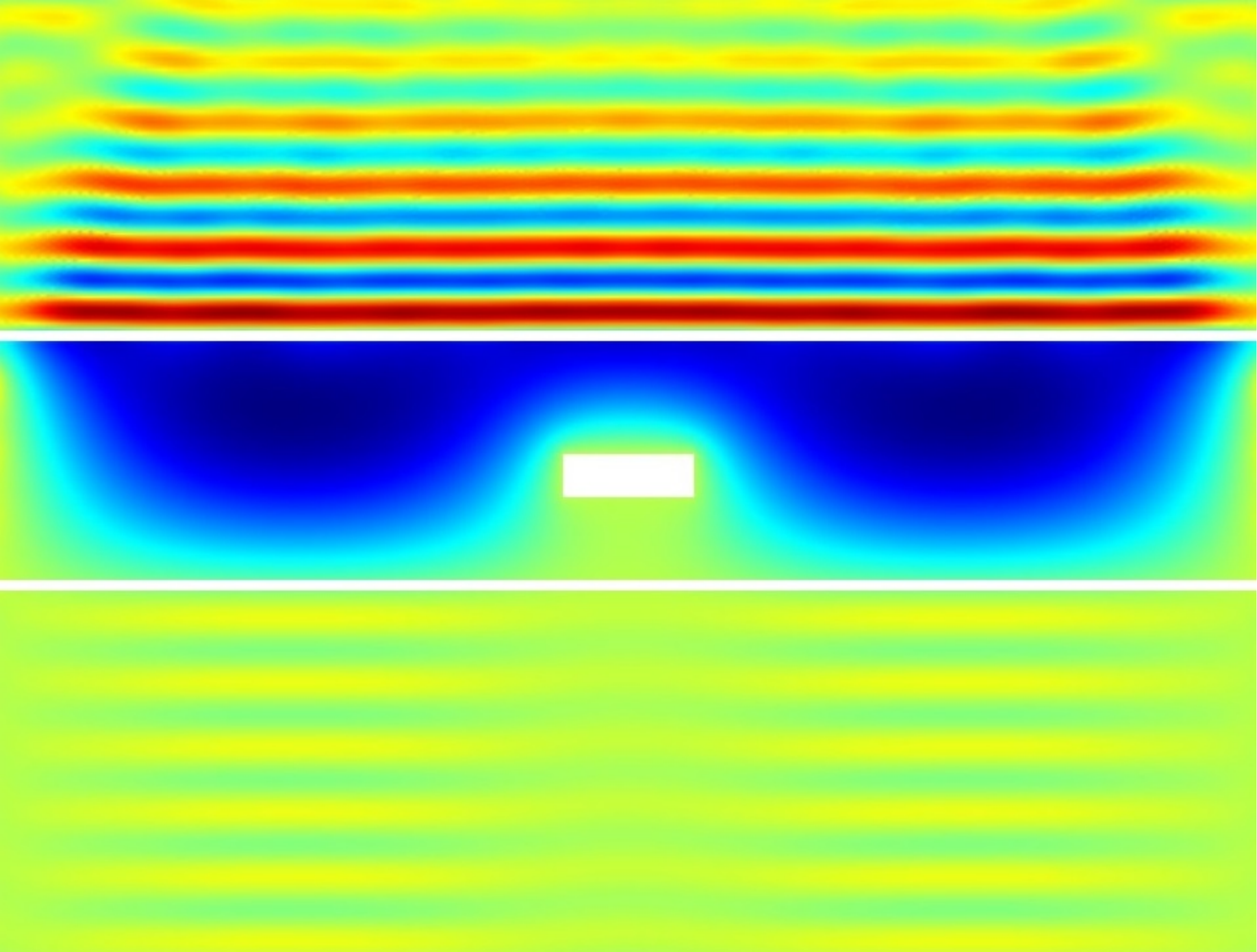} 
    \includegraphics[scale=0.188]{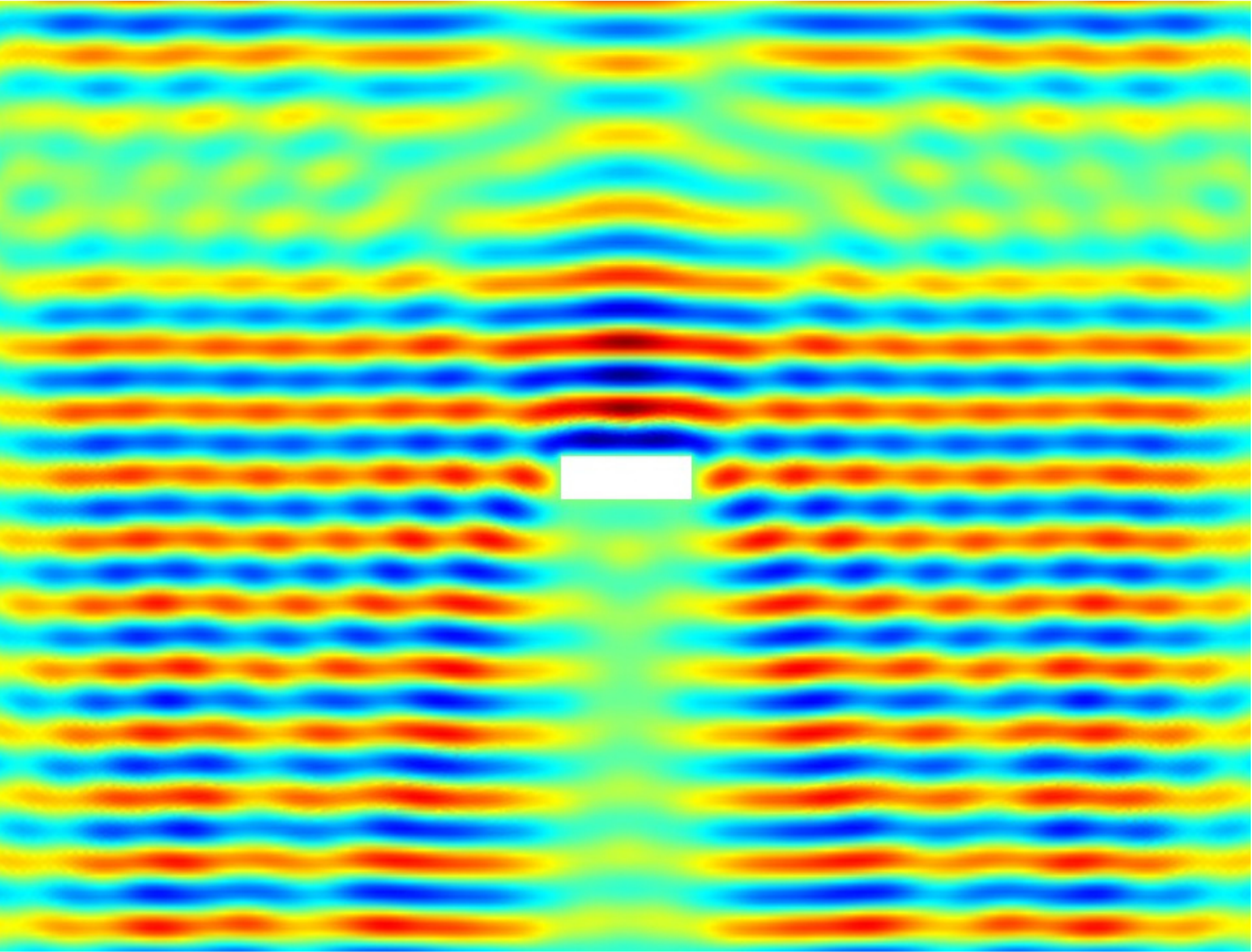}
\caption{Panel (a) shows nearly perfect transmission for a plane wave
at frequency $\Omega_0$ incident from above on an effective medium with properties computed by HFH for a frequency near $\Omega_0=2.215$ with PML on either sides of
the computational domain; A Plane wave
incident from above on the clamped obstacle at frequency
$\Omega_0$ for comparison.}
\label{fig:eff_cloak}
\end{figure}

\begin{figure}
\begin{center}
\includegraphics[scale=0.2]{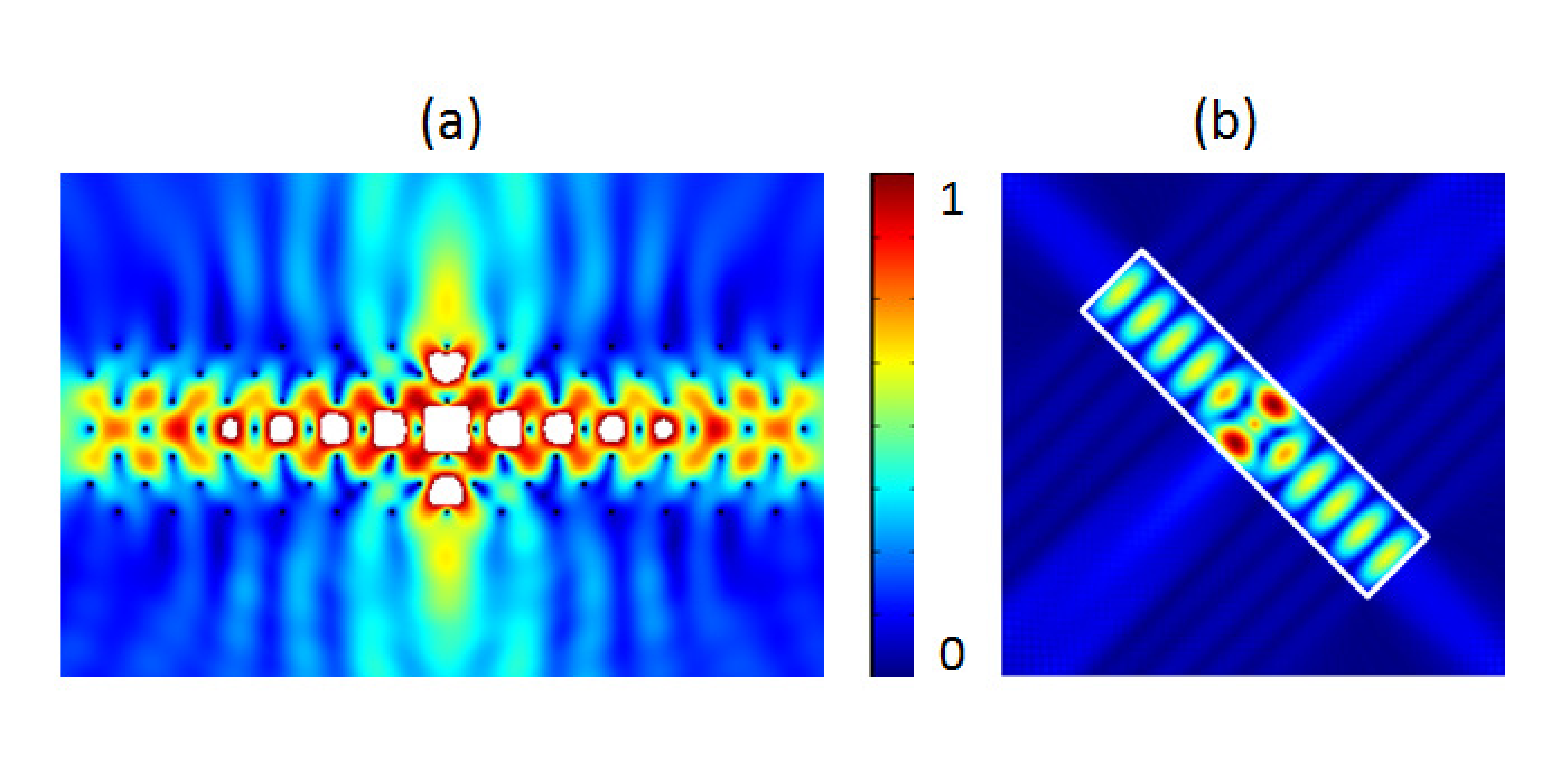}
\end{center}
\caption{Lensing effects in a PC (tilted array) of small Dirichlet holes:
(a) Lensing for a line source inside the PC at frequency $\Omega=2.7$; (b) Using equation (\ref{eq:fEff}) the PC is replaced by an effective medium in order to yield the same effect as in (a).}
\label{figseb3}
\end{figure}

\begin{figure}
\begin{center}
\includegraphics[scale=0.2]{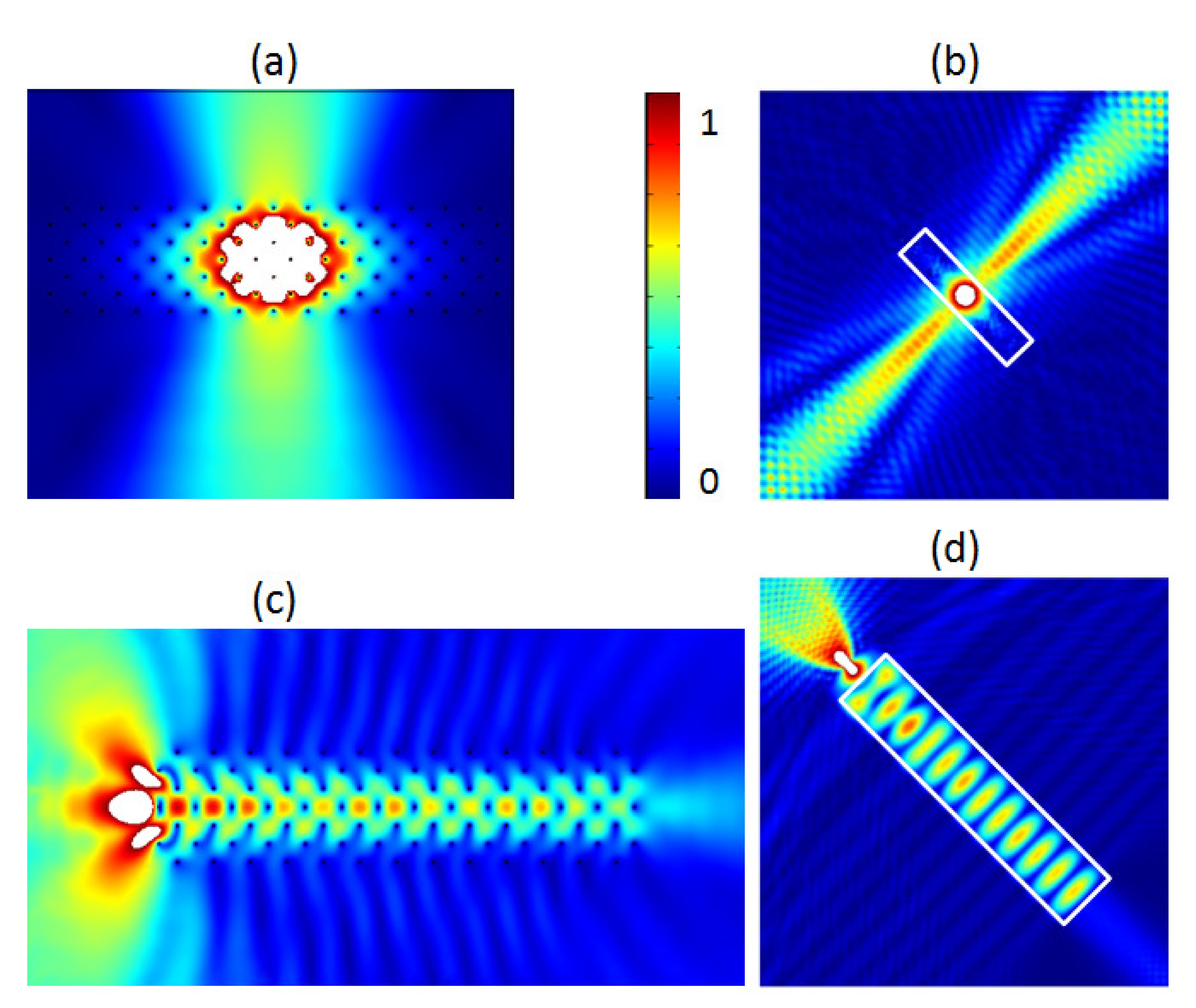}
\end{center}
\caption{Guiding and lensing effects in a PC (tilted array) of small Dirichlet holes:
(a) Omni-directive antenna for a line source inside the PC at
frequency $\Omega=\pi/2$; (b) The PC of (a) is replaced by an effective medium 
obtained by HFH with equation (\ref{eq:fEff}) in order to obtain the same physical effect;
(c) Endoscope for a line source
at $\Omega=2.7$; (d) The PC of (c) is replaced by an effective medium using equation (\ref{eq:effDirac}) to obtain the same physical effect.}
\label{fig10new}
\end{figure}

\subsection{Cloaking effects near Dirac-like cones}
\label{sec:cloak}
As noted in recent articles \cite{chan12a} a curious phenomenon occurs
in photonics near Dirac-like points which are the triple crossings shown
in, for instance, Fig. \ref{fig:r_0_01_zoom}. Let us consider the
lowest frequency triple point, near wavenumber $M$, shown in
Fig. \ref{fig:zero}; for the full finite element numerical simulations
we actually consider very small cylinders of radius $0.01$ in a square
array. The triple crossings occur because the dispersion relation of a
perfect medium (no cylinders) consists of light lines  folded in space
(one folded light line is shown in Fig. \ref{fig:zero}). The folded
light lines happen to satisfy the boundary condition for  the
infinitesimal point holes, and are perturbed slightly for finite
cylinders (as in Fig. \ref{fig:r_0_01_zoom}); the middle curve
passing between the Dirac-like cone is created by the inclusion and so
these effects co-exist. Importantly, this means that there is near
perfect transmission through an array of cylinders close to these triple
crossing frequencies as shown in Fig. \ref{figseb2}(a,b) as the
incoming plane wave, at $\Omega=2.215$,  does not see the inclusions. Inserting a large Dirichlet hole
within the array leads to virtually no reflection, the plane wave is
hardly reflected, and its transmission is nearly perfect
Fig. \ref{figseb2}(c) except for a
slight shadow zone in clear contrast to the uncloaked Dirichlet hole
Fig. \ref{figseb2}(d); 
If we remove the array of small holes, the defect is strongly
scattered by the wave. 

Replacing the array of holes with HFH is shown in
Fig. \ref{fig:eff_cloak} the quasiperiodic medium is replaced by a
homogeneous medium with the effective properties given by HFH. The
excitation is at a frequency $\Omega=2.12$, close to the standing wave
frequency $\Omega_0$. Inserting the values for $l=1$, $\Omega_0$,
$\Omega$ and $\beta=2/\pi^2$ into equation (\ref{eq:effDirac}) we
obtain the effective continuum equation,
\begin{equation}
{\hat f}_{0,x_ix_i}({\bf x})+0.0393{\hat f}_0({\bf x})=0.
\label{eq:cloakEff}
\end{equation}
Simulations show qualitatively the same effects, one has transmission
through the slab with plane waves incoming. 

\subsection{Lensing and wave guiding effects}
\label{sec:LensGuide}
Returning to the introduction we show in  Fig. \ref{fig:cross}(a,c) a
PC  composed of an array of $196$ holes with radius $0.4$; the
corresponding dispersion curves are shown in
Fig.  \ref{fig:2D_Dirichlet}. 

We now interpret these strongly anisotropic beams and how the HFH
represents this. In Fig \ref{fig:cross} (a,b)
 we choose frequency $\Omega=1.98$, which from Fig.
 \ref{fig:2D_Dirichlet} is near the standing wave frequency
 $\Omega_0=1.966$ at point $X(\pi/2,0)$. The reciprocal space is only
 shown for a portion of the irreducible Brillouin zone and one can use
 reflections of the triangle chosen to fill the entire square in the 
 Brillouin zone; due to these internal symmetries the effect is not
 only due to the first mode of Fig. \ref{fig:2D_Dirichlet} at point
 $X(0,\pi/2)$ of the reciprocal space, but also to the first mode, at
 point $G(\pi/2,0)$ (not shown). 
The effect is reproduced in Fig. \ref{fig:cross} (b) by combining two
effective medium equations, one for point $X$ and one for point $G$,
each one is responsible for a single diagonal, and the HFH equations are
\begin{align}
\quad -1.4778{\hat f}_{0,x_1x_1}({\bf x})+0.8837{\hat f}_{0,x_2x_2}({\bf x})+(\Omega^2-\Omega_0^2){\hat f}_0({\bf x})=0,
\\
\quad 0.8837{\hat f}_{0,x_1x_1}({\bf x})-1.4778{\hat f}_{0,x_2x_2}({\bf x})+(\Omega^2-\Omega_0^2){\hat f}_0({\bf x})=0,
\label{eq:EffCross}
\end{align}
with $(\Omega^2-\Omega_0^2)=0.0552$.
Note that these effective equations have opposite signs in the
coefficient, and so the governing equations are hyperbolic and not
elliptic, the result is that these direct energy along rays or
characteristics and the angle is given by the relative magnitude of
the $T_{11}$ and $T_{22}$; in this case roughly equal with the energy
directly along the diagonals. 

\begin{figure}
\begin{center}
\includegraphics[scale=0.5]{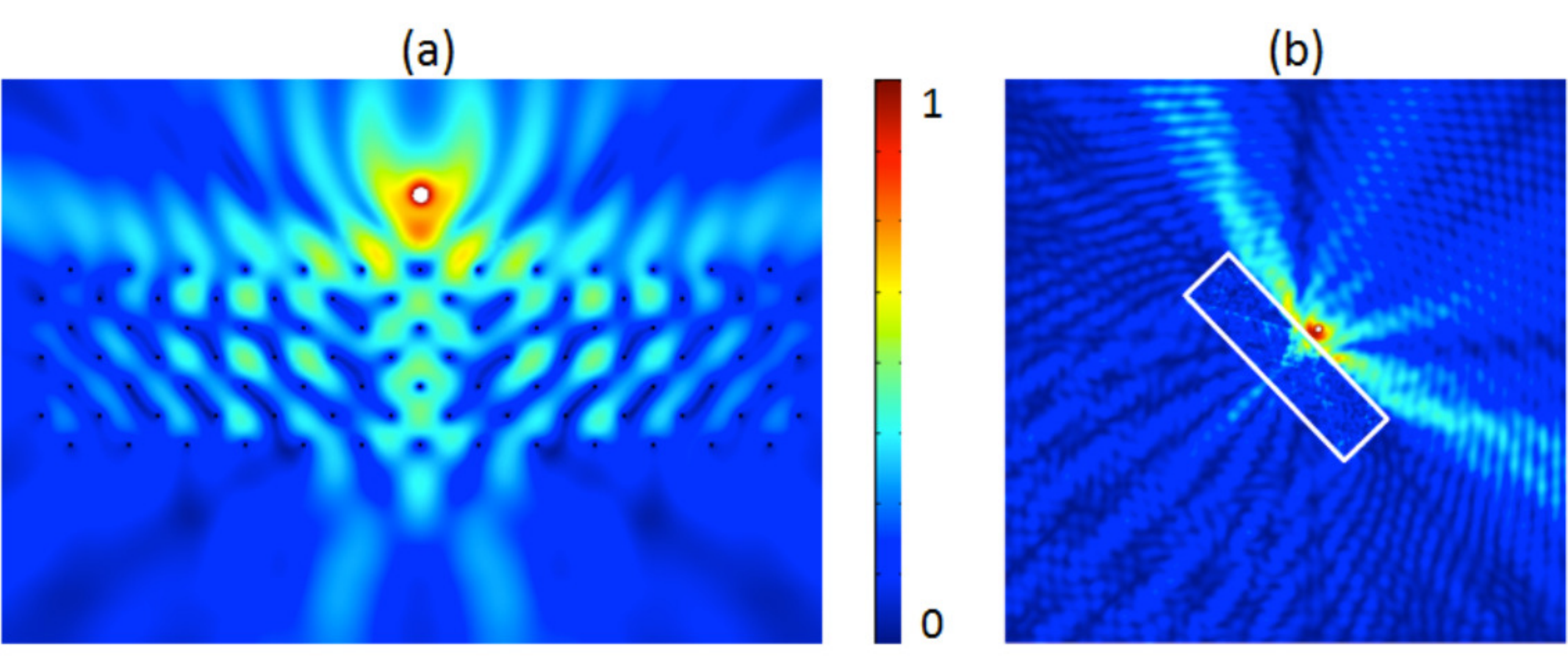}
\end{center}
\caption{Lensing effects in a PC (tilted array) of small Dirichlet holes acting as a flat lens near a frequency of $\pi/2$:
(a) A line source near the top of the PC yields an image on the other side. The PC is composed of $96$ very small holes arranged in a tilted square array with an angle of $\pi/4$. (b) shows clearly the same effect happening when the PC is replaced by a anisotropic effective medium obtained through HFH.}
\label{fig14}
\end{figure}

To illustrate this further in 
 panel (c) of Fig. \ref{fig:cross} we show a St George's cross effect obtained near the standing wave frequency $\Omega_0=2.744$. By homogenizing the PC and exciting it at a frequency of $\Omega=2.75$ we obtain two effective medium equations, one for $X$ and one for $G$ that are,
\begin{align}
\quad 3.1094{\hat f}_{0,x_1x_1}({\bf x})+0.085{\hat f}_{0,x_2x_2}({\bf x})+(\Omega^2-\Omega_0^2){\hat f}_0({\bf x})=0,
\label{eq:EffCrossReg1}
\\
\quad 0.085{\hat f}_{0,x_1x_1}({\bf x})+3.1094{\hat f}_{0,x_2x_2}({\bf x})+(\Omega^2-\Omega_0^2){\hat f}_0({\bf x})=0.
\label{eq:EffCrossReg2}
\end{align}
Each equation is responsible for a single line of the cross, the
orientation of this cross is explained by the almost zero coefficient,
in each effective equation, that does not permit propagation in the
vertical and horizontal direction for the respective equations
(\ref{eq:EffCrossReg1}) and (\ref{eq:EffCrossReg2}). The coefficient in front of ${\hat f}_0({\bf x})$ is equal to $0.033$. It is clear therefore that the strong anisotropy that is witnessed in
 full FE numerical simulations has its origins in the size and sign of
 the $T_{ij}$ coefficients from HFH and the effective equations can be
 used to gain quantitative understanding and predictions.

Another striking application involving wave propagation and an array
of small holes of radius $0.01$ is shown in Figs. \ref{figseb3}, \ref{fig10new} , \ref{fig14}
wherein we have tilted the array through an angle $\pi/4$. The
dispersion curves in Fig. \ref{fig:zero} are for holes of radius exactly
zero, but provide very good comparison for this small holes case.

We obtain a highly-directed emission at frequency $\Omega=1.6$ near
the standing wave frequency $\Omega_0=\pi/2$, shown in
Fig. \ref{fig10new}(a) with its effective medium counterpart in
\ref{fig10new}(b); in the dispersion curves this is the lowest
standing wave frequency at $X$. 
 There are two effective medium equations at this frequency which yield,
\begin{align}
\quad-11.42{\hat f}_{0,x_1x_1}({\bf x})+{\hat f}_{0,x_2x_2}({\bf x})+(\Omega^2-\Omega_0^2){\hat f}_0({\bf x})=0,
\\
\quad {\hat f}_{0,x_1x_1}({\bf x})-11.42{\hat f}_{0,x_2x_2}({\bf x})+(\Omega^2-\Omega_0^2){\hat f}_0({\bf x})=0,
\label{eq:effDirAnt}
\end{align}
where the difference of the frequency squares is equal to $0.0926$.
For forcing within the slab one again has a strongly anisotropic
response, the waves within the medium forming a cross-shape which, when it
strikes the edge of slab all gets radiated out into the surrounding
medium. Placing a source outside the slab, again at a frequency 
 near $\Omega_0=\pi/2$, Fig. \ref{fig14}(a) shows a PC that behaves
 like a flat lens. In panel (b) of that Fig. the effective medium
 yields the same effect and is also governed by two equations of the
 form of (\ref{eq:fEff}) with $T_{ij}$ coefficients equal to the ones
 of equation (\ref{eq:effDirAnt}). The frequency at which the lensing
 effect is obtained is $\Omega=1.56$ and so the coefficients in front
 of ${\hat f}_0({\bf x})$ are equal to $-0.0338$ in this case.

\begin{figure}
\begin{center}
\includegraphics[scale=0.2]{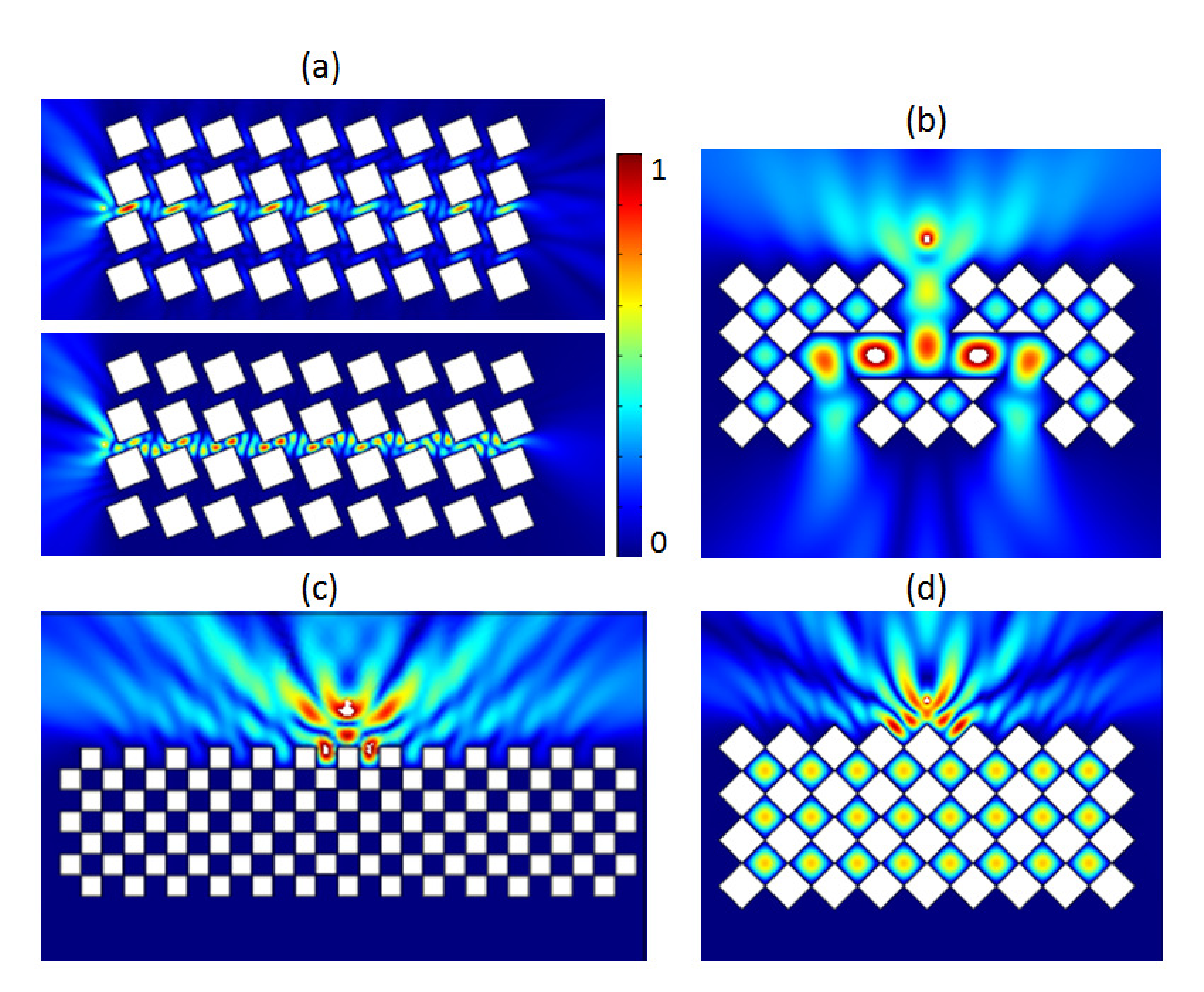}
\end{center}
\caption{Lensing and guiding effect through a PC composed of square Dirichlet holes which we vary their tilt angle:
(a) Lensing effects in a PC composed of Dirichlet square holes tilted by $\pi/8$ near the respective frequencies of $7.7$ and $8.25$ for the top and bottom figures (Fig \ref{tiltpi4pi8}a).
(b) The dispersion curves in Fig. \ref{tiltpi4pi8}b show the band
structure is composed of many band gaps separated by standing wave
modes, so that the structure acts as a perfect reflector for any
frequency in order to guide light. (c) shows a perfect reflector for a
source exciting the medium at frequency $4.95$, located in the second
band gap of Fig. \ref{tilt0}. (d) shows the same effect as in (c) for
a frequency of $5.0$ although in the present the filled parts of
material are excited but no energy escapes the PC and acts as an
energy trap.
}
\label{figTilt}
\end{figure}

Figs. \ref{figseb3}(a) and \ref{fig10new}(c) show lensing
effects where the PC is excited at $\Omega=2.7$, this frequency is
chosen as follows: When the array is rotated the relevant 
light line in the surrounding material, and its folding, emerge from
the point $M$ and are
 shown in Fig. \ref{fig:zero}
 we see that the energy from the source couples (as the light line
crosses the full dispersion curves at $\Omega=2.7$)  into one
of the linear Dirac-like cone modes that passes through
$\Omega_0=\pi$; this is the standing wave frequency of the Dirac-like cone
at $\Gamma$. We therefore model the effective medium with 
HFH using the analysis near $\Omega=\pi$, and it yields very similar results in Figs. \ref{figseb3}(b) and \ref{fig10new}(d), by homogenizing the PC with an effective material governed by equation,
\begin{equation}
{\hat f}_{0,x_ix_i}({\bf x})+0.3371{\hat f}_0({\bf x})=0,
\label{eq:lensingDirac}
\end{equation} 
obtained by simply replacing the source and standing wave frequencies together with $\beta=1/(2\pi^2)$ in equation (\ref{eq:effDirac}).

Finally, Fig. \ref{figTilt}(a) shows a unidirective PC for the first two flat modes of the band diagram in Fig. \ref{tiltpi4pi8}(a). The standing wave frequencies of the two flat modes in question are $\Omega_0=7.6$ and $\Omega_0=8.28$. The sources are excited near these frequencies at $\Omega=7.7$ and $\Omega=8.25$. HFH yields effective medium equations for each of the modes that are respectively,
\begin{equation}
\quad9.3649{\hat f}_{0,x_1x_1}({\bf x})+(\Omega^2-\Omega_0^2){\hat f}_0({\bf x})=0,
\label{eq:effTiltpi8a}
\end{equation}
and
\begin{equation}
\quad-45.8434{\hat f}_{0,x_1x_1}({\bf x})+(\Omega^2-\Omega_0^2){\hat f}_0({\bf x})=0,
\label{eq:effTiltpi8b}
\end{equation}
Equations (\ref{eq:effTiltpi8a}) and (\ref{eq:effTiltpi8b}) clearly show the unidirectivity of the two effective media. Similar equations with interchanged coeffcients stand, for the symmetric location of the Brillouin zone $G(0,\pi/2)$. 
Panels (b), (c) and (d) contain PCs with rotated square holes by an angle of $\pi/4$ with an unsual band diagram seen in Fig. \ref{tiltpi4pi8}(b). 
Panel (b) shows the guiding of a wave by means of perfect reflection for a line source excitated just outside the standing wave frequency of $\Omega=4.95$ while in panel (c) the same PC but this time rotated by an angle of $\pi/2$ acts as a perfect reflector. Panel (d) finally excites the medium right on the standing wave frequency of $\Omega=5$ and standing waves appear between the holes. Small gaps between the hole's edges make it possible for energy to pass from one cell to the other.  

\section{Concluding remarks}
\label{sec:conclude}
In this paper, we have presented a range of applications of the
high frequency homogenization theory, in the context of
zero-frequency stop band structures, which were previously thought to
be non-homogenizable: HFH has succeeded in capturing the finer
details of both dilute and densely packed photonic and
phononic crystals. Striking physical effects such as cloaking
via Dirac-like cones and directive antennas and
endoscope effects have been provided,
and fully explained, via HFH; importantly this is all shown to be
related to the potentially anisotropic $T_{ij}$ coefficients that
encode the local behaviour or their equivalent terms for multiple
crossings. 
The range of unsolved homogenization
problems in the wave community is
vast, and this new method now opens the door to efficient simulation
of multiscale structures. 

The current theory is limited to periodic, or nearly periodic media,
however one can envisage extensions in a variety of different
directions: In this paper we deal with photonic and phononic periodic
structures, but it should be possible to adapt our results
to quasi-crystals using a generalized form of the Floquet-Bloch
theorem in upper dimensional spaces.
Stochastic cases where the hole positions involve some random
perturbation, say,  might prove more arduous from a mathematical
standpoint, although a physicist might argue the period would be
replaced by the mean distance between the inclusions.  The key point
in order to relax the constraint of classical homogenization is the
assumption that the Floquet-Bloch theorem is applicable at least to
leading order.  Whenever this can be done, or an extension
of the Floquet-Bloch theorem can be used, HFH is applicable. Also of
interest are diffusion problems in composites, classic homogenization
is here again constrained by low frequencies (e.g. of a 
 periodic heat source), whereas much of the exciting physics lies
beyond the quasi-static regime again we hope that the insight provided
by HFH will lead  to progress in these directions.

\section*{Acknowledgements}
The authors thank Julius Kaplunov, Kirill Cherednichenko, Aleksey
Pichugin and Evgeniya Nolde for many enjoyable and useful
conversations. R.V.C. thanks the EPSRC (UK) for support through
research grant nunmber EP/J009636/1 and the University of
Aix-Marseille for support via a Visiting Professorship.  S.G. is
thankful for an ERC starting grant (ANAMORPHISM) that facilitates
collaboration with Imperial College London.

\section*{References}

\bibliographystyle{jphysicsB}
\bibliography{references}

\end{document}